\documentclass[fleqn,usenatbib]{mnras}

\usepackage{newtxtext,newtxmath}

\usepackage[T1]{fontenc}

\DeclareRobustCommand{\VAN}[3]{#2}
\let\VANthebibliography\thebibliography
\def\thebibliography{\DeclareRobustCommand{\VAN}[3]{##3}\VANthebibliography}

\usepackage{graphicx}	
\usepackage{amsmath}	
\usepackage[dvipsnames]{xcolor}

\DeclareMathAlphabet{\mathitbf}{OML}{cmm}{b}{it}
\newcommand{\mib}[1]{\ensuremath{\mathitbf{#1}}}

\title[Inverse Compton cooling in colliding-wind binaries]{Inverse Compton cooling of thermal plasma in colliding-wind binaries}

\author[J.~Mackey et al.]{
Jonathan Mackey$^{1}$\thanks{E-mail: jmackey@cp.dias.ie (JM)},
Thomas A.~K.~Jones$^{1,2}$,
Robert Brose$^{1}$,
Luca Grassitelli$^3$,
Brian Reville$^4$,
Arun Mathew$^{1}$
\\
$^{1}$Dublin Institute for Advanced Studies, Astronomy \& Astrophysics Section, DIAS Dunsink Observatory, Dublin, D15 XR2R, Ireland\\
$^{2}$School of Physics, Trinity College Dublin, The University of Dublin, Dublin 2, Ireland\\
$^{3}$Argelander-Institut für Astronomie, Auf dem Hügel 71, D-53121 Bonn, Germany\\
$^4$Max-Planck-Institut für Kernphysik, Saupfercheckweg 1, 69117 Heidelberg, Germany
}

\date{Accepted 13 September 2023. Received 18 August 2023; in original form 31 January 2023}

\pubyear{2023}

\begin{document}
\label{firstpage}
\pagerange{\pageref{firstpage}--\pageref{lastpage}}
\maketitle

\begin{abstract}
The inverse-Compton effect (IC) is a widely recognized cooling mechanism for both relativistic and thermal electrons in various astrophysical environments, including the intergalactic medium and X-ray emitting plasmas.
Its effect on thermal electrons is however frequently overlooked in theoretical and numerical models of colliding-wind binaries (CWB).
In this article, we provide a comprehensive investigation of the impact of IC cooling in CWBs, presenting general results for when the photon fields of the stars dominate the cooling of the thermal plasma and when shocks at the stagnation point are expected to be radiative.
Our analysis shows that IC cooling is the primary cooling process for the shocked-wind layer over a significant portion of the relevant parameter space, particularly in eccentric systems with large wind-momentum ratios, e.g., those containing a Wolf-Rayet and O-type star.
Using the binary system WR\,140 as a case study, we demonstrate that IC cooling leads to a strongly radiative shocked wind near periastron, which may otherwise remain adiabatic if only collisional cooling was considered.
Our results are further supported by 2D and 3D simulations of wind-wind collisions.
Specifically, 3D magnetohydrodynamic simulations of WR\,140 show a significant decrease in hard-X-ray emission around periastron, in agreement with observations but in contrast to equivalent simulations that omit IC cooling.
A novel method is proposed for constraining mass-loss rates of both stars in eccentric binaries where the wind-collision zone switches from adiabatic to radiative approaching periastron.
IC scattering is an important cooling process in the thermal plasma of CWBs.

\end{abstract}




\section{Introduction}

Colliding-wind binaries (CWB) are fascinating systems for studying massive stars, their strong radiation-driven winds, shock physics, particle acceleration and dust formation.
They emit across the electromagnetic spectrum from radio \citep{DouBeaCla05} to TeV gamma-rays \citep{HESS20_EtaCar}, predominantly non-thermal emission at the lowest and highest energies \citep{EicUso93}, and thermal radiation from the wind-wind collision region at X-ray energies \citep{Pol87, PolCorSte21}.
Hydrodynamic models and scaling relations for thermal X-ray emission were proposed by \citet{LuoMcCMac90}, who showed that the X-ray luminosity, $L_\mathrm{x}$, should scale with orbital separation, $d$, as $L_\mathrm{x}\propto d^{-1}$ for adiabatic shocked winds.
\citet{Uso92} obtained analytic solutions for the shape of the shocked region as a function of the properties of the two winds and derived the expected X-ray radiation.
\citet{SteBloPol92} investigated both adiabatic and radiative shocked layers and their dynamical instabilities with 2D simulations, discussing the implications for X-ray emission.
The X-ray emission from single and binary massive stars is reviewed by \citet{Rau22}.

The CWB systems, especially those with close or eccentric orbits, are of great interest.
They are not only the progenitors of high-mass X-ray binaries, but but also of supernovae and compact-object binary systems that can merge and become gravitational wave (GW) sources \citep[e.g.][]{LanSchSto20}.
The complex and poorly constrained post-main sequence evolution of massive stars \citep{Lan12, Smi14} is by far the great source of uncertainty on this path.
Mass-loss due to both the interaction between the constituent stars and the radiation-driven winds is key in determining the final masses, explosion taxonomy, orbital properties and, consequently, the rate of merger events across cosmic time \citep[e.g. ][ and references therein]{Vin22,EldSta22}.
For decades, the theoretical and observational challenges posed especially by the inhomogeneity of stellar winds hinder any solid prediction.
CWBs involving WR stars can present a great opportunity for independent constraints on some stellar parameters, such as the wind density and momentum ratio, just prior to the stellar demise.

Additionally, there has been substantial recent interest in the dust-producing subset of CWBs consisting of a carbon-rich Wolf-Rayet (WC) primary star in orbit with a less evolved companion \citep{LauEldHan20}, on account of their very high dust-formation rates.
3D hydrodynamical simulations of the WR\,140 \citep{EatPitVan22b} and WR\,104 \citep{SouLamMil23} systems including models of dust-grain growth have been presented to compare with existing observations and in anticipation of new results with JWST.
Dust shells in the WR\,140 system were recently observed using ground-based near-IR imaging \citep{HanTutLau22} and JWST mid-IR images \citep{LauHanHan22}.
The extreme CWB Apep was also observed at high angular resolution by \citet{HanTutLau20}, spatially resolving the dust plume.

Although dust production in WC binaries is poorly understood \citep{Che15}, a basic requirement is that the gas must cool below the dust sublimation temperature (typically about 1500\,K), and so must first cool from the coronal post-shock temperatures ($10^7-10^8$\,K) to nebular temperature ($\sim10^{4}$\,K) and then further to $\sim10^3$\,K.
This implies that the shocks must be radiative; \citet{SteBloPol92} provide approximate criteria for radiative shocks based on collisional cooling processes in the thermal plasma, namely X-ray/UV line emission and free-free (bremsstrahlung) emission.
These processes are routinely included in 2D and 3D hydrodynamic simulations \citep[e.g.][]{Pit09, PitPar10, MadGulOka13, LamMilLie17, EatPitVan22a, EatPitVan22b, SouLamMil23}.

Inverse-Compton (IC) scattering of background radiation fields by non-relativistic electrons is well known as a cooling mechanism for hot, low-density plasmas, from the thermal Sunyaev-Zeldovich effect \citep{SunZel72, Bir99} in Galaxy Clusters, and from models of the thermal evolution of the intergalactic medium \citep[e.g.][]{WieSchSmi09}.
\citet{Kin03} recognised that IC cooling may also be important in models of AGN-driven outflows from galaxies, although the equations used are only applicable to relativistic electrons.
\citet{FauQua12} and \citet{RicFau18} calculated results for IC cooling from both relativistic and non-relativistic electrons in AGN-driven outflows, also including the case where electrons and ions have different temperatures.
The process is also included by \citet{HopWetKer18} in the FIRE-2 radiative cooling model for cooling of hot gas off the CMB.

In CWBs the circumstellar medium experiences an intense radiation field with a spectral energy distribution that approximately follows that of a blackbody with the temperature of the surface of the nearest star, $T_\mathrm{eff}$.
\citet{Che76} first argued that IC cooling of the thermal electrons could play an important role in CWBs, predicting that for short-period systems the X-ray emission could be suppressed by up to a factor of 20 due to IC cooling.
This was further discussed in \citet{WhiChe95}, who noted that \citet{SteBloPol92} overpredicted the thermal X-ray emission from V444\,Cyg in a hydrodynamical model, possibly because they neglected IC cooling.
\citet{MyaZhe93} included IC cooling in models of CWBs that included wind acceleration, finding that it can be important for determining X-ray emission from close binaries.
The only other implementation of IC cooling in hydrodynamical simulations (as far as we are aware) is by \citet{StLMofMar05}, who used 2D simulations to model the wind collision in the SMC CWB Sanduleak 1 (a WO4+O4 system).
The process is also briefly mentioned, but not implemented, in \citet{LamDubLes12}.

These results serve as motivation to investigate IC cooling in close binary systems with and without radiative shocks, especially because the process is rarely included in theoretical modelling, simulations or interpretation of X-ray observations.
In this paper we make a more comprehensive study of IC cooling of shocked stellar wind in CWB systems and demonstrate an implementation in multi-dimensional hydrodynamic simulations.
Section~\ref{sec:theory} introduces the equations describing gas properties in CWB systems as well as estimates for the escape time, IC and free-free cooling times of the plasma, and details of the implementation in a magnetohydrodynamic code.
Section~\ref{sec:general-res} derives some general results for binary systems containing a Wolf-Rayet (WR) star and a companion with a weaker wind.
These results are then applied in section~\ref{sec:wr140} to the specific case of the CWB system WR\,140 (located in Cygnus), first with analytic estimates, then 2D and 3D simulations.
Some discussion is presented in section~\ref{sec:discussion}, including a proposal of a method to better constrain the mass-loss rate in the weaker component of the binary system, and conclusions in section~\ref{sec:conclusions}.

\section{Theoretical model and numerical methods}
\label{sec:theory}

\subsection{Colliding-wind binaries}

The shocks in a CWB form where the ram pressure of the two winds are equal, forming a contact discontinuity and two reflected shocks that quickly reach a stationary configuration in the absence of radiative cooling and dynamical instabilities \citep{LuoMcCMac90, SteBloPol92}.
The winds are usually substantially ionized by the EUV radiation from the stars and so we can use an adiabatic index $\gamma=5/3$.
We consider the expanding wind to accelerate according to a beta-law with $\beta=1$.
This is a rough approximation \citep[e.g.][]{SanVin20} but is sufficient for our purposes.
The wind velocity, $V(r)$, has a dependence on radial distance from the centre of the star, $r$, of
\begin{equation}
V(r) = V_\infty\left( 1 - \frac{R_\star}{r}\right) \;,
\end{equation}
where $R_\star$ is the stellar surface or some equivalent surface where the wind velocity is small.
The winds are supersonic by definition, and usually hypersonic so that, at the wind termination shock, we can use the Rankine-Hugoniot jump conditions for a strong shock.
The post-shock density, $\rho_\mathrm{ps}$ is therefore simply obtained from the wind density, $\rho_w(r)$, as
\begin{equation}
\rho_\mathrm{ps} = 4\rho_w(r) = \frac{\dot{M}}{\pi r^2 V(r)} \;,
\end{equation}
where $\dot{M}$ is the mass-loss rate of the star, assumed spherically symmetric and smooth.

The location of the wind collision can be calculated as a function of the wind-momentum ratios of the two stars \citep{LuoMcCMac90, MiySugMae22}:
\begin{equation}
\eta \equiv \frac{\dot{M}_1 V_{1}(d_1)}{\dot{M}_2 V_{2}(d_2)} = \left( \frac{d_1}{d_2} \right)^2 \;,
\end{equation}
where $d_1$ and $d_2$ are the distances from the centres of stars 1 and 2 to the wind-collision zone, respectively.
If one ignores the wind acceleration and uses $V_{\infty}$ for the velocities then this is trivial to solve for a given total separation $d=d_1+d_2$.
The situation is much more complicated including the wind acceleration because within the acceleration zone of star 1, the wind of star 2 will be decelerated by the same radiation force that accelerates the wind of star 1.
For the purposes of calculating the shock position we ignore the wind acceleration.

We define the mean mass per electron, $\mu_\mathrm{e}$, in units of the proton mass, $m_\mathrm{p}$, as
\begin{equation}
\mu_\mathrm{e} \equiv \frac{\rho}{n_\mathrm{e} m_\mathrm{p}} = \left( X + \frac{1}{2}[Y+Z] \right)^{-1} \;,
\end{equation}
where $n_\mathrm{e}$ is the electron number density and the elemental composition of the gas is defined by $[X,Y,Z]$, being the mass fraction of H, He and heavier elements, respectively.
The mean mass per particle, $\mu$, is as usual
\begin{equation}
\mu \equiv \frac{\rho}{(n_\mathrm{e}+n_\mathrm{ion}) m_\mathrm{p}} = \left( 2X + \frac{3Y}{4}+\frac{Z}{2} \right)^{-1} \;,
\end{equation}
where $n_\mathrm{ion}$ is the ion number density.
These relations result if the gas is fully ionized and if we approximate that $Q/A \approx 0.5$ and $(1+Q) / A \approx0.5$ for metals with atomic number $Q$ and mass number $A$.
For standard Galactic ISM abundances we obtain $\mu_\mathrm{e} = 1.18$ and $\mu=0.62$.

The internal energy of the gas is
\begin{equation}
E_\mathrm{int} = \frac{3}{2} (n_\mathrm{e}+n_\mathrm{ion}) k_\mathrm{B} T = \frac{3\rho}{2\mu m_\mathrm{p}}k_\mathrm{B} T \;,
\end{equation}
where $k_\mathrm{B}$ is the Boltzmann constant.
The postshock temperature (in the limit that a single fluid temperature is applicable) is
\begin{equation}
T_\mathrm{ps} = \frac{3\mu m_\mathrm{p}}{16 k_\mathrm{B}} v_\mathrm{sh}^2 \;,
\end{equation}
where $v_\mathrm{sh}$ is the shock velocity.
If we assume the shock structures are relaxed to their equilibrium positions, then $v_\mathrm{sh} = V(r)$, the wind velocity at the shock location.

From this, the postshock adiabatic sound speed can be calculated assuming a strong shock that decelerates the gas by $4\times$:
\begin{equation}
c_\mathrm{s} \equiv \sqrt{\frac{\gamma k_\mathrm{B} T_\mathrm{ps}}{\mu m_\mathrm{p}}} = \frac{\sqrt{5}v_\mathrm{sh}}{4} \;.
\end{equation}
This is used to calculate the advection time for gas to leave the wind-collision region,
\begin{equation}
\tau_\mathrm{adv}\equiv \frac{d_2}{c_\mathrm{s}} = \frac{4d_2}{\sqrt{5}v_\mathrm{sh}} \;,
\end{equation}
where $d_2$ is the separation from the contact discontinuity to the centre of the star with the weaker wind. 
For a CWB system with $\eta<1$ the shock will form a bow shape around the star with the weaker wind, and so the distance to this star should be used for the timescale.
This is later used to compare with the cooling timescales to determine at what separations a shock will be radiative or adiabatic (\citealt{SteBloPol92} refer to this as $t_\mathrm{esc}$).

\citet{SteBloPol92} introduced the cooling parameter
\begin{equation}
\chi\equiv \frac{\tau_\mathrm{cool}}{\tau_\mathrm{adv}} \approx \frac{v_8^4 d_{12}}{\dot{M}_{-7}} \;,
\label{eqn:chi}
\end{equation}
where $v_8$ is the wind velocity in units of $10^8$\,cm\,s$^{-1}$, $d_{12}$ is the distance from the star to the contact discontinuity in units of $10^{12}$\,cm and $\dot{M}_{-7}$ is the mass-loss rate of the star in units of $10^{-7}$\,M$_\odot$\,yr$^{-1}$.
This is based on the assumption that the cooling rate is approximately constant with temperature in the temperature range of shocks with $v_8\sim1$.
The relation has been very widely used in the literature since its introduction.
If IC cooling is important (for which $\tau_\mathrm{ic}$ has no dependence on temperature or density) then this relation will not be applicable.

\subsection{IC cooling of the thermal plasma}
\label{sec:ic-theory}

The IC heating rate of photons (and cooling rate of the gas) for a non-relativistic gas is given by \citep[e.g.][]{RybLig79}
\begin{equation}
\dot{E}_\mathrm{ic} = \frac{4 k_\mathrm{B} T}{m_\mathrm{e} c^2}
                      \sigma_\mathrm{T} c  n_\mathrm{e} U_\gamma \;(\mathrm{erg\,cm^{-3}\,s^{-1}})\;,
\label{eqn:iccool}
\end{equation}
where $m_\mathrm{e}$ the electron mass, $T$ the gas temperature, $\sigma_\mathrm{T}$ the Thomson cross-section, and the photon energy density is $U_\gamma$.
Note that effects due to the anisotropy of the radiation field scale with $\beta = v/c$ and so does not arise here because we are considering non-relativistic electrons.
The above expression holds if $T$ is much larger than the radiation temperature of the photon field, but goes to zero if the temperatures are comparable.
For the numerical implementation we replace $T$ with $(T-T_\mathrm{eff})$, where $T_\mathrm{eff}$ is the surface temperature of the star.

The IC cooling time of the thermal plasma (assuming electrons and ions have a single temperature, $T$; see section~\ref{sec:general-res}) is $\tau_\mathrm{ic}\equiv E/\dot{E}_\mathrm{ic}$, which depends only on $U_\gamma$ and can be expressed as
\begin{equation}
\tau_\mathrm{ic} = \frac{3 \mu_\mathrm{e} m_\mathrm{e} c}{8 \mu \sigma_\mathrm{T}} (U_\gamma)^{-1} \;.
\end{equation}
A similar expression was obtained by \citet{MyaZhe93}.
This can be expressed according to parameters of the system as
\begin{equation}
\tau_\mathrm{ic} = 2.89\times10^5 \,\mathrm{s} \;
                      \left(\frac{L_\star}{10^6 \mathrm{L}_\odot}\right)^{-1}
                      \left(\frac{r}{10^{13}\,\mathrm{cm}}\right)^2\;,
\label{eqn:tau-comp}
\end{equation}
where $r$ is the distance from the star (with luminosity $L_\star$) to the shocked gas and we have used ISM abundances for $\mu$ and $\mu_\mathrm{e}$.
This expression assumes that $r$ is at least a few stellar radii, so that the dilution factor reduces to the inverse square law.
\citet{Kin03} used the relativistic expression for the energy loss rate and so obtained that $\tau_\mathrm{ic}$ scaled inversely with the energy of the gas (or $\propto v^{-2}$ for shock velocity $v$ in the expanding AGN outflow).
This may be true for outflows from AGN if they are driven by the pressure of relativistic particles, but it is not appropriate for CWBs.
It is important here that $\tau_\mathrm{ic}$ is independent of gas temperature because CWBs with large eccentricity frequently have wind-wind collisions within the wind-acceleration region where the shock velocity (and post-shock temperature) is not as simple to estimate as for wider binaries where the terminal velocity of the wind can be used.

For IC cooling we can define a parameter similar to the $\chi$ of \citet{SteBloPol92} as
\begin{equation}
\chi_\mathrm{ic} \equiv \frac{\tau_\mathrm{ic}}{\tau_\mathrm{adv}} 
 = \frac{1.61\, d_{12} v_8}{L_5} \;,
\end{equation}
where $L_5$ is the luminosity of the star in units of $10^5\, \mathrm{L}_\odot$, $d_{12}$ and $v_8$ have the same meaning as in Equation~\ref{eqn:chi}.
When $\chi_\mathrm{ic}<1$ then we expect a shock to be radiative.

The Compton y parameter, $y_\mathrm{c}$, measures the fractional energy change per photon passing through a region, and for the non-relativistic limit is given by
\begin{equation}
y_\mathrm{c} = \frac{4k_\mathrm{B} T}{m_\mathrm{e} c^2} \max (\tau_\mathrm{es},\tau_\mathrm{es}^2)  \;,
\end{equation}
where $\tau_\mathrm{es} = \int n_\mathrm{e} \sigma_\mathrm{T} d\ell$ is the electron-scattering optical depth of the region being considered.
If $y_\mathrm{c}$ is small then the spectrum and total energy of the photon field are not significantly modified by IC scattering \citep[e.g.][]{RybLig79}.
The effect of IC scattering on the thermal electron population, on the other hand, depends also on the photon flux in the region: the electrons may be efficiently cooled even if $y_\mathrm{c}\ll1$, if the photon field is sufficiently intense.
The local cooling rate $\dot{E}_\mathrm{ic}$ is the photon energy flux, $cU_\gamma$, times the integrand of the $y_\mathrm{c}$ integral along the photon trajectory.

Equation~\ref{eqn:iccool} only holds in the limit that $y_\mathrm{c}\ll1$ (otherwise the full Kompaneets equation should be solved), so we should estimate the typical values expected in CWBs.
\citet{LuoMcCMac90} calculate that the thickness of the shocked region, $\Delta r$, in an adiabatic wind-wind collision is $\Delta r \approx r/8$, where $r$ is the distance to the star.
We can approximate this region as having constant density and pressure, with
$\rho = 4\rho_w(r) = \dot{M}/ (\pi r^2 v_\mathrm{sh})$ as above.
Then the integral becomes trivial and we obtain (for $\tau_\mathrm{es}<\tau_\mathrm{es}^2$):
\begin{equation}
  y_\mathrm{c} = \frac{4k_\mathrm{B} T}{m_\mathrm{e} c^2} \frac{\dot{M} \sigma_\mathrm{T}}{\pi r v_\mathrm{sh} \mu_\mathrm{e} m_\mathrm{p}}
  \frac{\Delta r}{r}   \approx 8.0\times10^{-6} \frac{\dot{M}_{-7} v_8}{d_{12}}  \;.
\end{equation}
Even for dense and fast winds with shocks close to the stellar surface, $y_\mathrm{c}\ll1$ will always be satisfied and the medium is optically thin.
This simplifies the numerical implementation in multi-dimensional simulations because the radiation field is not significantly affected by IC scattering.
IC cooling can therefore be implemented as a purely local process depending only on the gas properties and distance from the two stars in the binary system.

\subsection{Free-free cooling of the thermal plasma}

For stellar winds from classical WR and O stars (with $V_\infty \gtrsim 2000$\,km\,s$^{-1}$), the postshock temperature is large enough that free-free cooling (bremsstrahlung) is the dominant collisional cooling process.
All ions are fully ionized by the shock and so there are no spectral lines available for cooling.

Bremsstrahlung (free-free radiation) has a stronger scaling with density but weaker scaling with temperature than IC cooling:
\begin{equation}
\dot{E}_\mathrm{ff} \approx 1.68\times10^{-27} n_\mathrm{e} n_i Q_i^2 \sqrt{T} \;(\mathrm{erg\,cm^{-3}\,s^{-1}})\;,
\end{equation}
where $n_i$ and $Q_i$ are the number density and atomic number for each element $i$ (with velocity-averaged Gaunt factor $=1.2$, \citealt{RybLig79}).
Summing over all elements gives
\begin{equation}
\dot{E}_\mathrm{ff} \approx 1.68\times10^{-27} \left(\frac{\rho}{m_\mathrm{p}}\right)^2 \sqrt{T} \frac{X + Y + 0.5 \sum_{i>2} X_i Q_i}{\mu_\mathrm{e}} \;(\mathrm{erg\,cm^{-3}\,s^{-1}})\;,
\end{equation}
where $X_i$ is the mass fraction of element $i$, and again we assume the atomic mass number $A_i=2Q_i$ for $i>2$.
For ISM abundances the last term is $\approx1$.
The cooling time from free-free radiation, $\tau_\mathrm{ff}$ is then given by
\begin{equation}
\tau_\mathrm{ff} = \frac{3 \mu_\mathrm{e} m_\mathrm{p} k_\mathrm{B}}{3.36\times10^{-27} \mu \left(X+Y+0.5\sum_{i>2} X_iQ_i\right)} \frac{\sqrt{T}}{\rho} \;.
\end{equation}
This can be expressed in terms of parameters of the system (for the ISM abundances quoted above) as
\begin{equation}
\tau_\mathrm{ff} = 5.02\times10^6 \,\mathrm{s}
    \left(\frac{\dot{M}}{10^{-6}\, \mathrm{M}_\odot\,\mathrm{yr}^{-1}}\right)^{-1}
    \left(\frac{r}{10^{13}\,\mathrm{cm}}\right)^2
    \left(\frac{V_\infty}{2\times10^{8}\,\mathrm{cm\,s}^{-1}}\right)^2
\label{eqn:tau-ff}
\end{equation}
where $r$ is the distance from the star to the shocked gas.
Comparing Equations~(\ref{eqn:tau-comp}) and~(\ref{eqn:tau-ff}) we can expect that there are regions of parameter space where IC cooling will dominate.
Similarly to the IC cooling calculation above, we can define for free-free cooling:
\begin{equation}
\chi_\mathrm{ff} \equiv \frac{\tau_\mathrm{ff}}{\tau_\mathrm{adv}} 
 = \frac{7.02\, d_{12}v_8^3}{\dot{M}_{-7}} \;,
\end{equation}
where $d_{12}$, $v_8$ and $\dot{M}_{-7}$ have the same meanings as in Equation~\ref{eqn:chi}.
The scaling is different from that of \citet{SteBloPol92} because they considered lower temperature gas with $v_8\approx 1$ where $\dot{E}$ is approximately flat with temperature, whereas we assume $v_8\gtrsim1.5$  where free-free cooling dominates.

For the 2D and 3D simulations presented later we use the collisional ionization-equilibrium cooling function of \citet{EatPitVan22a}, with lookup tables kindly provided to us by J.~Pittard.
The tabulated cooling rate is $\dot{E}_\mathrm{cie}  = (\rho/\mu)^2 \Lambda_\mathrm{cie}(T)$, and the cooling timescale for collisional processes is $\tau_\mathrm{cie} \equiv E_\mathrm{int} / \dot{E}_\mathrm{cie}$.
This cooling function includes free-free radiation, which dominates at high temperatures, but also includes metal-line cooling at lower temperatures and so is not simply a power-law in temperature like $\dot{E}_\mathrm{ff}$.

\begin{figure}
\centering
  \includegraphics[width=0.95\columnwidth]{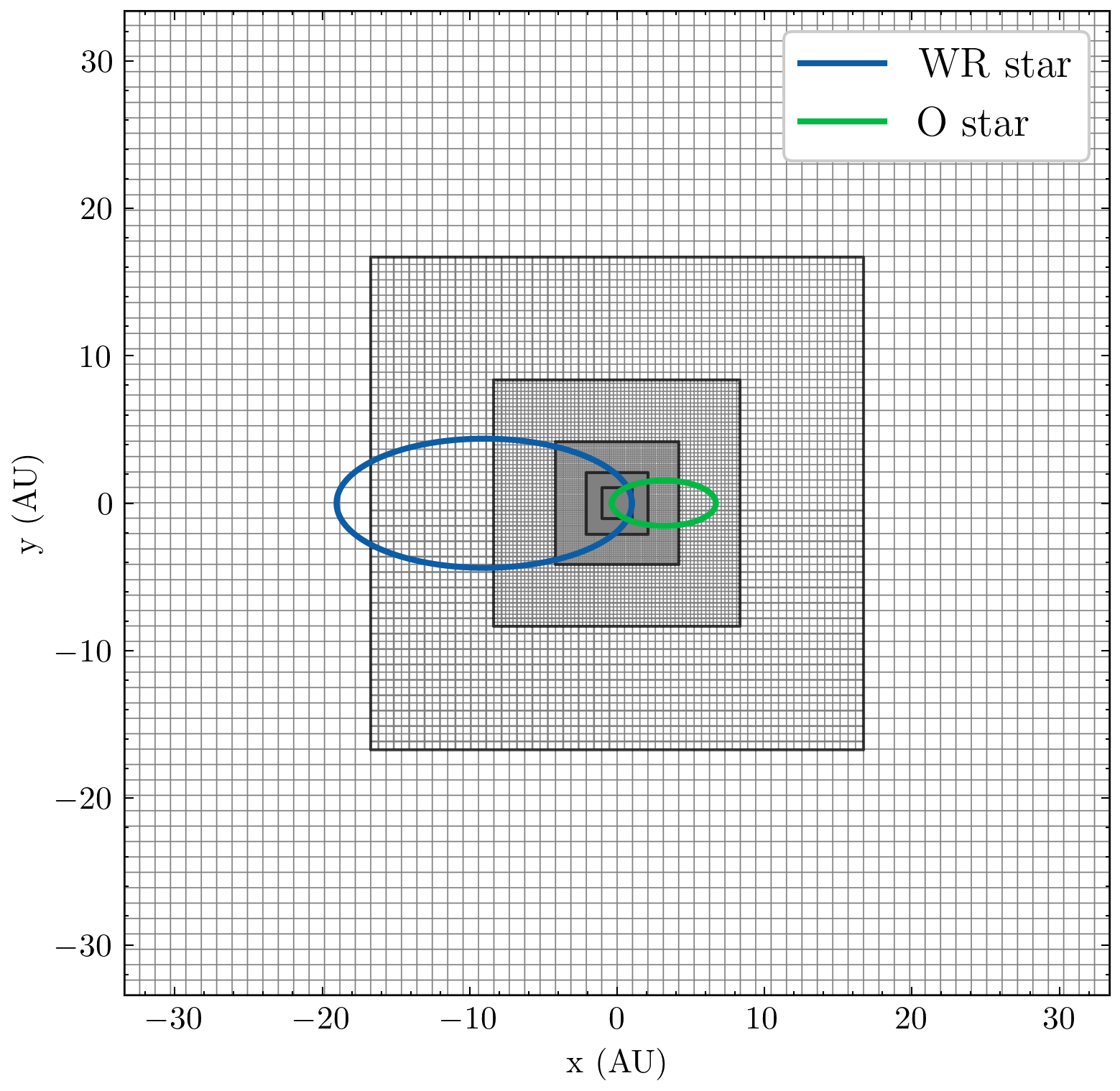}
  \caption{
    Example of the nested grids for the midplane of a 6-level simulation with $64\times64$ cells in each level and a domain width of $10^{15}$\,cm (about 67\,au).
    The side length of each grid is a factor of 2 smaller than the next coarsest level with equal numbers of cells in each level.
    The 3D simulations in section~\ref{sec:wr140-3d} have an extra coarse level $2\times$ larger; this is omitted so that the finest level can be easily seen.
    The orbital solution for WR\,140 is superimposed on the grid, with the system's centre of mass at the origin.
  }
  \label{fig:grid}
\end{figure}

\subsection{Implementation of a numerical algorithm}
\label{sec:sim-methods}

We used the (magneto-)hydrodynamics code \textsc{pion} \citep{MacGreMou21} to make 2D and 3D simulations of the wind-wind collision of the two stars in the WR\,140 system for different stellar separations.
The 2D hydrodynamic (HD) simulations use cylindrical coordinates in the $R$-$z$ plane (with rotational symmetry in the $\theta$ direction).
The two stars are static and on the $z$-axis (i.e., orbital motion is ignored), and static mesh-refinement (with multiply nested grids) with 6 levels is used to focus resolution on the wind-collision region.
The 3D magnetohydrodynamic (MHD) simulations use Cartesian coordinates centred on the centre of mass of the system, with the two stars moving along their orbit from the initial position through periastron.
Again static mesh-refinement is used as demonstrated in \citet{MacGreMou21}, very similar to the setup of \citet{EatPitVan22a}, here with 7 levels of refinement.
An example of a slice through the grid midplane is plotted in Fig.~\ref{fig:grid} with the orbits of the two stars in WR\,140 overlaid (see section~\ref{sec:wr140-3d} for more details).

Four code enhancements were required to enable the 2D and 3D simulations: high-resolution and robust Riemann solvers, wind acceleration, stellar orbital motion and composition-dependent radiative cooling.
The HLLD and Roe solvers typically used for MHD and HD proved not sufficiently robust for the CWB simulations, even with the switch to HLL solvers for cells with strong shocks \citep{MigZanTze12}.
An extra criterion was added to revert to HLL solver if the density jump from one cell to the next exceeded a factor of 5 for MHD or 10 for HD (i.e., strong contact discontinuity).
This limits how sharply a contact discontinuity can be resolved, but still allows the development of dynamical instabilities at the interface.

A simple model for wind acceleration is implemented, using a $\beta$ law with $\beta=1$ and an initial wind velocity of $V_0=4 c_\mathrm{s} (T_\mathrm{eff})$ at $r=R_\star$:
\begin{equation}
V(r) = V_0 + (V_\infty-V_0)\left( 1 - \frac{R_\star}{r}\right) \;,
\end{equation}
The gas at every point within $r<100R_\star$ is given a radial acceleration such that a freely expanding wind will follow the $\beta$-law velocity profile.
The exception to this is hot shocked gas which lacks the opacity required to accelerate the wind, for which we linearly decrease the acceleration from its nominal value to zero in the temperature range $[1-2]\times10^6$\,K.
We do not distinguish between wind composition, so that the WR wind is decelerated by the radiation field of the O star.
This is a relatively crude approximation compared with implementations of the so-called CAK coefficients \citep[e.g.][]{Pit09}, but it is sufficient to demonstrate the importance of IC cooling.
More detailed wind acceleration will be implemented in future work.

For 3D simulations we implemented orbital motion using a leapfrog integrator to solve the 2-body gravitational problem for unequal point masses.
The stars are given initial positions and velocities, appropriately pre-calculated for the period, eccentricity and semi-major axis of the system and at the desired orbital phase.
The wind speeds are always much larger than the orbital velocity (by about a factor of 10), and so the hydrodynamic timestep is always more restrictive than that of the orbital motion.
The interior of the star is given unphysical values so that it is easily identified on plots (low and constant values of density, pressure, velocity and magnetic field).
The wind boundary region is defined as a sphere whose radius, $r_\mathrm{w}$, satisfies the criteria:
\begin{enumerate}
\item $r_\mathrm{w}$ at least 2 cells larger than the stellar radius, $R_\star$, on the coarsest level that intersects the boundary region;
\item $r_\mathrm{w}$ at least 7 cells on the coarsest level that intersects the boundary region; and
\item for MHD simulations, $r_\mathrm{w}$ at least 2 cells larger than the Alfv\'en radius of the wind on the coarsest level that intersects the boundary region.
\end{enumerate}
All cells with distance from the star of $r\leq r_\mathrm{w}$ are labelled as not part of the simulation domain, and so their properties are not evolved in time except insofar as the star moves across the grid.
The wind boundary cells are updated when the star moves by a distance $0.1\Delta x$ on the finest grid level, where $\Delta x$ is the cell diameter.
These criteria ensure that the wind is supersonic and super-Alfv\'enic on entering the simulation domain, but it imposes limits on how close to the star we can resolve the hydrodynamics and it imposes an upper limit on the surface magnetic field strength of the stars.
Neither of these limitations is a problem for this paper because we are not interested here in strongly magnetised stars.

Radiative cooling follows the two-component cooling function of \citet{EatPitVan22a}. 
We distinguish between the two wind compositions by use of a passive scalar and the total cooling rate is a linear combination of the two cooling functions, with weights determined by the passive scalar.
We do not include the dust cooling of \citet{EatPitVan22a}, but instead add the IC cooling from both stars as described above.
We assume the gas is optically thin to the bulk of the stellar continuum radiation, and so do not need to employ ray-tracing to calculate attenuation factors.
Instead we calculate the local radiation energy density at every point on the domain for each star, according to an integration of the specific intensity, $I(\Omega)$, over solid angle:
\begin{equation}
U_\gamma \equiv \frac{1}{c} \int I(\Omega) d\Omega  = \frac{L_\star}{2\pi R_\star^2 c} \left(1 - \sqrt{1-\left(\frac{R_\star}{r}\right)^2} \right) \;,
\end{equation}
where $r$ is the distance from the centre of the star to the cell in question.
This includes the so-called \textit{dilution factor} that takes account of the finite angular size of the star.
This is stored as an extra scalar field on each grid for each star.
The IC cooling rate is then calculated from the sum of the energy densities.

Following the description of \textsc{PION} in \citet{MacGreMou21}, the full set of equations that are solved are
\begin{align}
\frac{\partial\rho}{\partial t}  + \nabla \cdot (\rho \mib{v}) &= \dot{m}_\mathrm{w} \;, \\
\frac{\partial\rho\mib{v}}{\partial t}  + \nabla \cdot (\mib{v}\otimes \rho \mib{v} - \mib{B}\otimes\mib{B})  &=  \nonumber \\
  -\nabla \left(p + \frac{\mib{B}^2}{2}\right) & +\rho(\mib{g}-\mib{g}_\mathrm{rad}) \textcolor{blue}{-(\nabla\cdot\mib{B})\mib{B}} \;,  \\
\frac{\partial E}{\partial t}  + \nabla \cdot ( [E+p]\mib{v} -\mib{B}[\mib{v}\cdot\mib{B}] 
   & + \textcolor{olive}{c_\mathrm{h} \psi \mib{B}} ) = \nonumber \\
  \frac{\rho^2}{m_\mathrm{p}^2} (\Gamma - \Lambda) + \dot{E}_\mathrm{ic} +  
  \rho\mib{v}\cdot(\mib{g}-\mib{g}_\mathrm{rad}) & 
  \textcolor{blue}{- (\nabla\cdot\mib{B})(\mib{v}\cdot\mib{B})} 
  \textcolor{olive}{- (\nabla\psi)\cdot(\mib{v}\psi)} \;, \\
\frac{\partial \mib{B}}{\partial t} - \nabla \times (\mib{v}\times\mib{B}) - c_\mathrm{h} \nabla \psi &= 
  \textcolor{blue}{-(\nabla\cdot\mib{B})\mib{v}}  \;, \nonumber \\
\frac{\partial \psi}{\partial t} + c_\mathrm{h}\nabla \cdot \mib{B} &=  \textcolor{olive}{-(\nabla\psi)\cdot\mib{v}} \;, \\
\end{align}
where $p$ is the gas pressure, $\rho$ the density, $\mib{v}$ the velocity, \mib{B} the magnetic field,
\begin{equation}
E = \frac{1}{2}\rho\mib{v}^2 + \frac{p}{\gamma-1} + \frac{1}{2}\mib{B}^2 + \textcolor{olive}{\frac{1}{2}\psi^2}
\end{equation}
is the total energy density of the plasma with adiabatic index $\gamma=5/3$, $\dot{m}_\mathrm{w}$ the source term of mass for stellar winds leaving the star, $\psi$ is the scalar field from the GLM-MHD method for divergence cleaning \citep{DedKemKro02, DerWinGas18} and $c_\mathrm{h}$ is the hyperbolic wavespeed associated with this field.
The net force of gravity and radiation pressure, $\rho(\mib{g}-\mib{g}_\mathrm{rad})$, is chosen to give a $\beta$-law acceleration profile in the freely expanding wind of each star, as described above.
The RHS terms multiplying $\nabla\cdot\mib{B}$, written in blue, are the \citet{PowRoeLin99} source terms, and the terms involving $\psi$, written in olive colour, follow \citet{DerWinGas18} reformulation of the GLM-MHD equations.

The cooling function $\Lambda$ is given by \citet{EatPitVan22a} as described above, and the photoheating rate $\Gamma$ is calculated by assuming there are sufficient ionizing photons to keep the plasma at least singly ionized, with each recombination resulting in a subsequent photoionization and liberation of the excess photon energy as heat \citep{GreMacHaw19}.
IC cooling (or heating) is calculated as described above, assuming the $y_\mathrm{c}\ll1$ limit (justified above and in the following section).
The coupling of radiation to matter is therefore only through optically-thin radiative cooling or heating of the plasma through IC scattering, and through the radiation force that we impose such that it reproduces a $\beta$-law wind acceleration.
We are not solving the equations of radiation hydrodynamics; rather we make reasonable approximations and assumptions that allow us to estimate the impact of the radiation on the thermal and dynamical state of the gas.

\begin{figure}
\centering
  \includegraphics[width=\columnwidth]{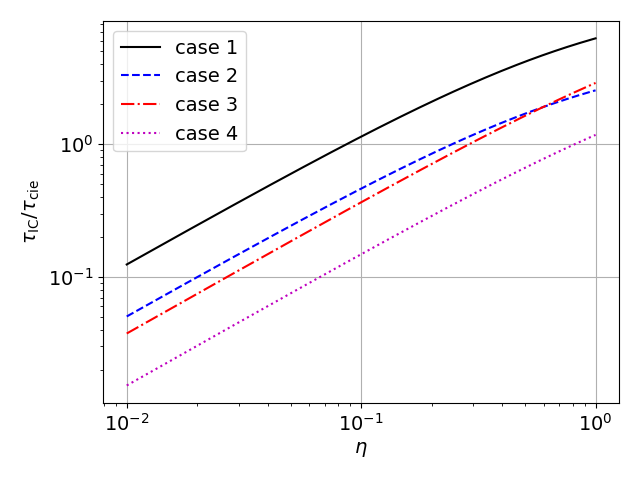}
  \caption{Ratio of $\tau_\mathrm{ic}/\tau_\mathrm{cie}$ for the four cases considered in Section~\ref{sec:general-res} as a function of the wind momentum ratio, $\eta$.
  }
  \label{fig:comp-ff}
\end{figure}

\begin{figure*}
\centering
  \includegraphics[width=0.95\columnwidth]{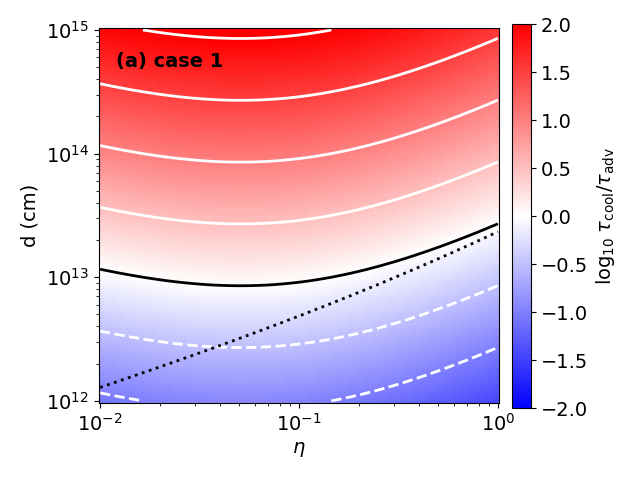}
  \includegraphics[width=0.95\columnwidth]{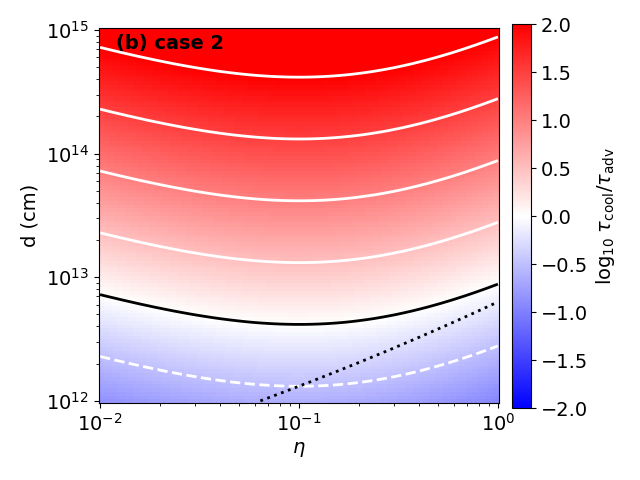} \\
  \includegraphics[width=0.95\columnwidth]{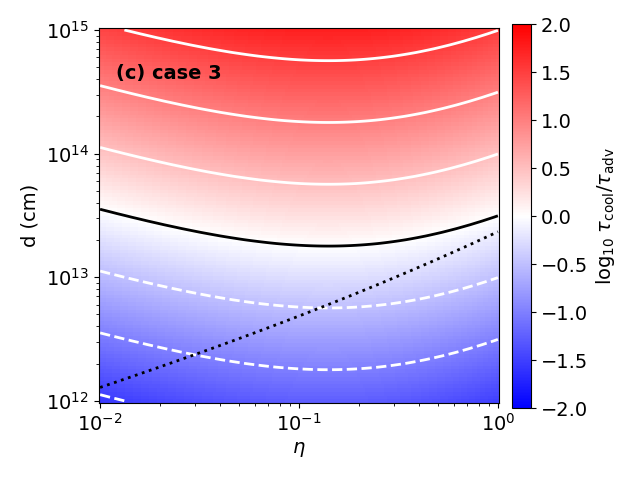}
  \includegraphics[width=0.95\columnwidth]{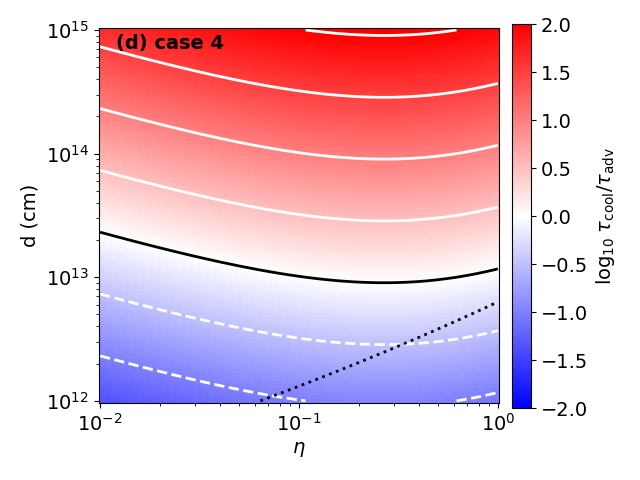}
  \caption{Ratio $\log_{10} \tau_\mathrm{cool}/\tau_\mathrm{adv}$ for the 4 cases considered in Section~\ref{sec:general-res} as a function of the wind momentum ratio, $\eta$, and star separation, $d$, plotted on the colour scale and the white contours.
  Contours are plotted at intervals of 0.5, with negative contours dashed and the zero contour the solid black line.
  The dotted black line shows the same zero contour when IC cooling is not included.
  }
  \label{fig:tcool-tadv}
\end{figure*}

\section{General results}
\label{sec:general-res}
Here we consider the impact of IC cooling from the combined radiation fields of two stars in a close binary system on the shocked gas from each wind, as a function of the orbital separation of the two stars.
We use the wind momentum ratio $\eta = \dot{M}_2 V_{\infty,2} / \dot{M}_1 V_{\infty,1}$ to estimate the shock location following \citet{SteBloPol92}.
If the shock is radiative then the shocked region is thin and the shocks effectively coincide with the contact discontinuity of the two winds.
Then we can calculate the IC cooling associated with the radiation from both stars, as a function of orbital separation, $d$, again assuming the shock is far enough from both stars that the inverse-square law gives an accurate estimate of the energy density (including the dilution factor only increases the energy density and reduces the IC cooling time by less than a factor of 2).

Star 1 is considered to have parameters similar to a WR star, with $L_1 = 3\times10^5\,$L$_\odot$ and $\dot{M}_1 = 3\times10^{-5}$\,M$_\odot$\,yr$^{-1}$.
For star 2 we consider four cases:
\begin{enumerate}
\item Case 1: $L_2 = 3\times10^5\,$L$_\odot$, $v_\infty=2000$\,km\,s$^{-1}$;
\item Case 2: $L_2 = 3\times10^5\,$L$_\odot$, $v_\infty=3000$\,km\,s$^{-1}$;
\item Case 3: $L_2 = 10^6\,$L$_\odot$, $v_\infty=2000$\,km\,s$^{-1}$; and
\item Case 4: $L_2 = 10^6\,$L$_\odot$, $v_\infty=3000$\,km\,s$^{-1}$.
\end{enumerate}
For simplicity we take $v_\infty$  of stars 1 and 2 to be the same, so that the wind densities are equal at the wind-collision region (because of ram-pressure balance).
The mass-loss rate of star 2 is a variable in the following analysis, defined by $\dot{M}_2 = \eta \dot{M}_1$ for equal wind velocities.
We consider a range of values for $d$ and $\eta$, namely $d\in[10^{12},10^{15}]$\,cm ($\approx$0.067-67\,au) and $\eta\in[0.01,1]$.
This covers most of the parameter space of binary systems where we expect to encounter radiative shocks.

Note that $\tau_\mathrm{ic}$, $\tau_\mathrm{ff}$ and $\tau_\mathrm{cie}$ all have the same scaling with distance from the star, because the photon density and wind mass-density both follow an inverse-square law.
We plot the ratio $\tau_\mathrm{ic}/\tau_\mathrm{cie}$ in Fig.~\ref{fig:comp-ff} for the four cases above, showing that for most values of $\eta$ the ratio is $<1$ and IC cooling dominates over collisional cooling.
For Case 1, with the densest wind and lowest-luminosity star 2, collisional cooling is dominant for $\eta>0.1$ but, for the cases where star 2 is $3\times$ more luminous, IC cooling dominates almost up to $\eta=1$.
The curves are almost indentical if we replace $\tau_\mathrm{cie}$ with $\tau_\mathrm{ff}$ because of the high post-shock temperature of the plasma.

The total cooling time is
\begin{equation}
\tau_\mathrm{cool} = 
\frac{E_\mathrm{int}}{\dot{E}_\mathrm{cie} + \dot{E}_\mathrm{ic,1} + \dot{E}_\mathrm{ic,2}} =
\left(\frac{1}{\tau_\mathrm{cie}}+  \frac{1}{\tau_\mathrm{ic,1}}+\frac{1}{\tau_\mathrm{ic,2}}\right)^{-1},
\end{equation}
where subscripts 1 and 2 refer to IC cooling from the two radiation fields.
Note that this cooling prescription is identical to the one used in the multi-dimensional simulations in the following section.

In Fig.~\ref{fig:tcool-tadv} we plot $\log_{10} \tau_\mathrm{cool}/\tau_\mathrm{adv}$ for the range of parameter space in $d$ and $\eta$ and for the 4 cases.
All of the area below the black solid contour (blue region) is radiative, and the area above (red region) is is effectively adiabatic.
We see that only the close binaries are expected to have radiative shocks (or eccentric binaries near periastron), and the parameter space with $\eta\ll1$ is dominated by IC cooling whereas the upturn in the contours as $\eta \rightarrow 1$ is driven by free-free cooling (shown by the cyan contours).
It is clear that estimates of whether a shock is radiative will be wrong both qualitatively and quantitatively if IC cooling is not included.
For $\eta\ll1$, the range of $d$ where $\tau_\mathrm{cool}<\tau_\mathrm{adv}$ can be up to $50\times$ larger when IC cooling is included (Fig.~\ref{fig:tcool-tadv}).

\begin{figure}
\centering
  \includegraphics[width=\columnwidth]{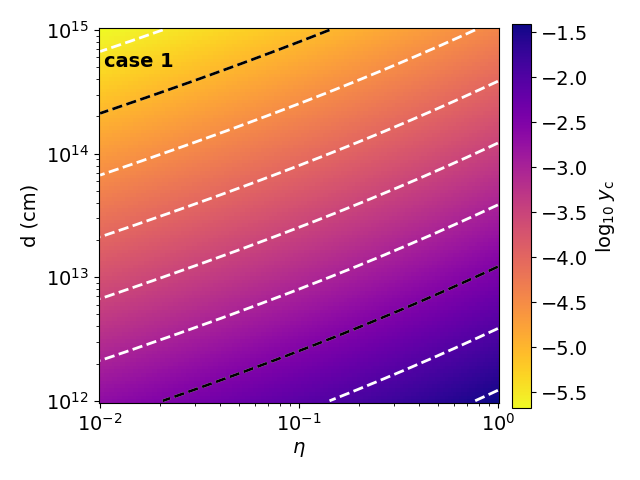}
  \caption{
    Compton y parameter, $y_\mathrm{c}$ for case 1 in Section~\ref{sec:general-res} as a function of the wind momentum ratio, $\eta$, and star separation, $d$, plotted on the colour scale and with white contours.
  Contours are plotted at intervals of 0.5 dex, with black contours at -5.0 and -2.5.
  }
  \label{fig:compy}
\end{figure}

The Compton y parameter of this parameter space is plotted in Fig.~\ref{fig:compy} for Case 1.
The other 3 cases are similar, with $y_\mathrm{c}$ within a factor of 3 of this plot.
For almost all of the parameter space $y_\mathrm{c}\ll1$, reaching a maximum of $y_\mathrm{c}\approx0.03$ at the bottom right corner of the plot, i.e., close binaries with similar mass-loss rates.
As shown in Section~\ref{sec:ic-theory}, $y_\mathrm{c} \propto 1/d$
In order to reach $y_\mathrm{c}\sim1$, a binary system would need to have two stars with Wolf-Rayet type winds (if not stronger) and very small separations (orbital period on the order of a few days or less).
For most of the known CWB systems we are safe in assuming $y_\mathrm{c}\ll1$.

\begin{figure*}
\centering
  \includegraphics[width=0.95\columnwidth]{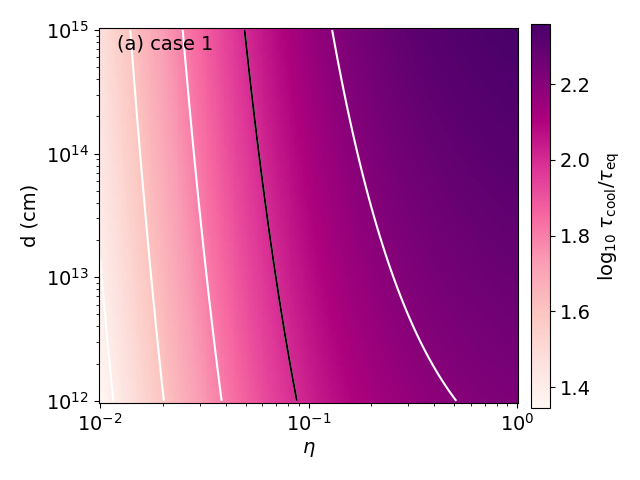}
  \includegraphics[width=0.95\columnwidth]{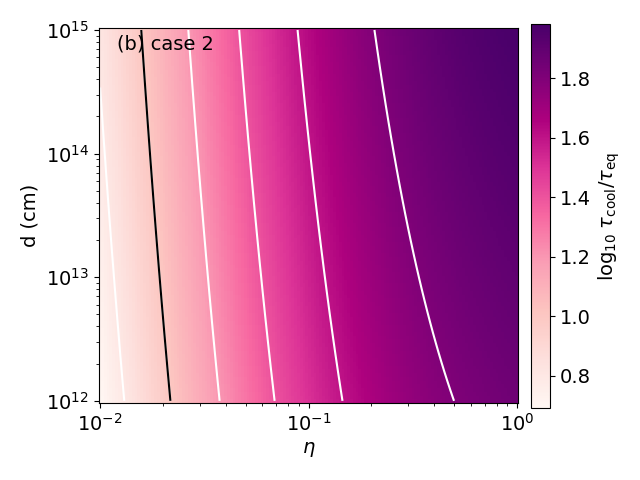} \\
  \includegraphics[width=0.95\columnwidth]{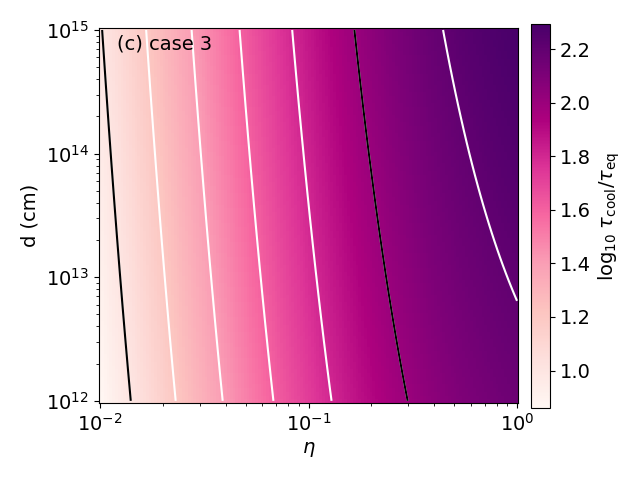}
  \includegraphics[width=0.95\columnwidth]{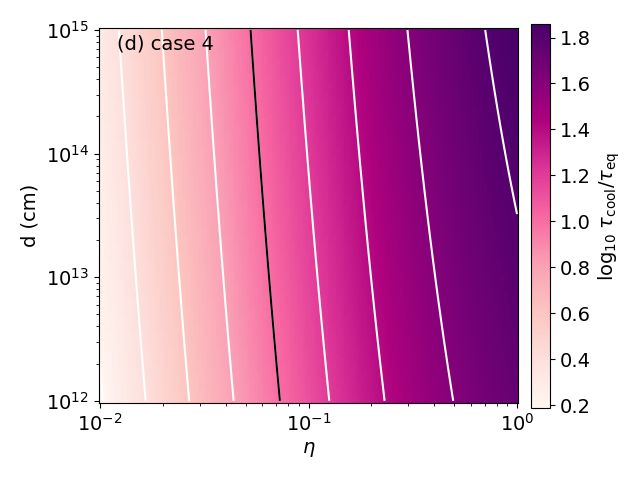}
  \caption{Ratio $\log_{10} \tau_\mathrm{cool}/\tau_\mathrm{eq}$ for the 4 cases considered in Section~\ref{sec:general-res} as a function of the wind momentum ratio, $\eta$, and star separation, $d$, plotted on the colour scale and with white contours.
  Contours are plotted at intervals of 0.2 dex, with black contours at 1.0 and 2.0.
  }
  \label{fig:teq}
\end{figure*}

The immediate postshock temperature of the ions and electrons is not the same, but they relax to the same temperature via Coulomb interactions on a timescale calculated by \citet{Spi62}.
\citet{FauQua12} argue that IC cooling can be less efficient in AGN-driven outflows than the above equations suggest because of inefficient electron-ion coupling.
Using the equilibration timescale, $\tau_\mathrm{eq}$, quoted in, e.g., \citet{PolCorSte05}, we plot in Fig.~\ref{fig:teq} the ratio $\log_{10} \tau_\mathrm{cool}/\tau_\mathrm{eq}$.
This shows that $\tau_\mathrm{cool}/\tau_\mathrm{eq} > 1$ for all cases over the whole parameter space, and for most of the parameter space the ratio is $\sim[10-100]$, and so the electrons should reach the equilibrium single-fluid, post-shock temperature before they begin to cool.
The region of parameter space where radiative shocks occur due to IC cooling (the lower left part of each panel) is, however, the region with the lowest $\tau_\mathrm{cool}/\tau_\mathrm{eq}$ ratio, and so it could be that for stars with exceptionally fast winds and heavy element compositions, the ratio could approach unity and the single-fluid assumption would then no longer be valid.
For example case 4 in Fig.~\ref{fig:teq} has the strongest photon field with large wind velocity, and here $\tau_\mathrm{cool}/\tau_\mathrm{eq} \rightarrow 1.6$ at the bottom left corner of the panel (small separation and extreme wind momentum-ratio).
In this case we do not expect full electron-ion equilibrium to be attained, and the IC cooling rate is limited by the collisional electron-heating rate.

These results show that IC cooling is a crucial ingredient in determining whether or not shocks in CWB systems will be radiative or adiabatic, and it is especially important for systems with very disparate mass-loss rates ($\eta\ll1$) and where the stars come very close to each other for at least part of their orbit ($d\lesssim3\times 10^{13}$\,cm, or 2.0\,au).
One system that satisfies both of these requirements is WR\,140, discussed in detail in the next section.

\section{WR 140 as a case study for IC cooling}
\label{sec:wr140}
WR\,140 is one of the archetypal CWBs, exhibiting time-dependent X-ray \citep{PolCorSte21, MiySugMae22} and radio \citep{DouBeaCla05} emission with a 7.9 year period.
The system is highly eccentric ($e = 0.8993$; \citealt{ThoRicEld21}), and produces large quantities of dust near periastron \citep{WilMarMar09,LauHanHan22}.
\citet{DouBeaCla05} spatially resolved the synchrotron emission from the shocked wind with radio interferometry, and this was further interpreted by \citet{PitDou06} in the context of an analytic model of particle acceleration and radiation.
High-energy $\gamma$-ray emission has not been detected \citep{Psh16}, although the bright diffuse emission from the Cygnus region makes any detection very challenging.
Numerical modelling with 2D \citep{ZheSki00} and 3D \citep{RusCorOka11} adiabatic simulations has given some insight into the X-ray lightcurve, and a semianalytic model by \citet{Zhe21} has highlighted the important role of absorption in determining the observed lightcurve.
\citet{PolCorSte21} have modelled the absorption as a function of orbital phase and constructed an intrinsic hard-X-ray lightcurve from the wind-collision region.

For CWB systems the composition of the two winds may be very different, and this affects the cooling timescales and postshock gas temperature.
The O star can be assumed to have a Galactic elemental abundance, with mass fractions of H, He, and metals (Z) given by $[X,Y,Z]\approx[0.7,0.28,0.02]$.
For the WC primary star the wind contains essentially no H and is strongly enriched in He and C, so we take $[X,Y,Z]=[0.0,0.55,0.45]$ \citep{EatPitVan22b}.
For the above abundances we obtain the usual results for the O star: $\mu=0.62$ and $\mu_\mathrm{e}=1.18$, and for the WR star $\mu=1.57$ and $\mu_\mathrm{e}=2.0$.
For free-free emission from the shocked WC wind we can assume that essentially all of the metals are C \citep{EatPitVan22b}.
Using these differing abundances, and given that the wind velocities of the two stars are not identical, the cooling and advection timescales for the two shocks are not the same (as was assumed in the previous section).

\begin{table}
\centering
\caption{Table of binary, stellar and wind properties for the two components of the CWB system WR\,140.  References for the values are given in the text.}
\label{tab:wr140}
\begin{tabular}{lcc} 
\hline
\textbf{Property} & \textbf{O star} & \textbf{WC star} \\
\hline
$L_\star$ (L$_\odot$)     & $10^6$              & $10^{5.5}$         \\
$R_\star$ (R$_\odot$)     & 26.0                & 2.16  \\
$M_\star$ (M$_\odot$)     & 29.3 & 10.3 \\
$V_\infty$ (km\,s$^{-1}$) & 3100                & 2860                \\
$\dot{M}$ (M$_\odot$\,yr$^{-1}$) (low-$\eta$)  & $0.87\times10^{-6}$ & $4.3\times10^{-5}$ \\
$\dot{M}$ (M$_\odot$\,yr$^{-1}$) (mid-$\eta$)  & $0.9\times10^{-6}$ & $2.2\times10^{-5}$ \\
\hline
                          & \textbf{binary system} &   \\
\hline
Semimajor axis, $a$   & $2.1\times10^{14}$\,cm & (14.0\,au) \\
eccentricity, $e$         & 0.8993        &   \\
periastron separation, $d_\mathrm{p}$ & $2.1\times10^{13}$\,cm & (1.4\,au) \\
apastron separation, $d_\mathrm{a}$ & $4.0\times10^{14}$\,cm  & (26.7\,au) \\
\end{tabular}
\end{table}

\subsection{Properties of the binary system}
The properties of the two stars in the system and their winds are somewhat uncertain.
For IC cooling the key parameter is the luminosity of each star.
\citet{DouBeaCla05} estimate $\log L/L_\odot = 6.18$ for the O-star secondary, and $\log L/L_\odot = 5.5$ for the WC-type Wolf-Rayet (WR) primary.
The WR star is more evolved but has currently a lower mass than the secondary because its envelope has been stripped via stellar wind and/or binary mass-transfer.
\citet{WilMarMar09} estimate $\log L/L_\odot = 5.93$ and 5.5 for the two stars.
We will take $\log L_\mathrm{O}/L_\odot = 6.0$ for the O star and $\log L_\mathrm{WR}/L_\odot = 5.5$ for the WR star.
The radiation temperature is not important for the IC cooling, as long as the typical photon energy is much smaller than the energy of the shocked electrons.
For wind speed of $2000$\,km\,s$^{-1}$ (which we may take as a minimum expected shock speed given the terminal velocities quoted below) the post-shock temperature is $T_\mathrm{e}=5.6\times10^7$\,K, much larger than the effective temperature of any star.

The stellar radius is important for wind acceleration, and we take $R_\mathrm{O}=26\,$R$_\odot$ for the O star \citep{WilMarMar09} and $R_\mathrm{WR}=2.16\,$R$_\odot$ (or $1.5\times10^{11}$\,cm) for the WR star.
The relevant radius for wind acceleration of the WR star is the sonic point, and~\citet{GraLanGri18} showed that this is at approximately [0.9-1.5]\,R$_\odot$ for the classical WR stars (the exact value is not important for any of the calculations in this paper).
The terminal wind velocity of both stars is quite well constrained, with $V_{\infty,\mathrm{O}}=3100$\,km\,s$^{-1}$ for the O star and $V_{\infty,\mathrm{W}}=2860$\,km\,s$^{-1}$ for the WR star \citep{WilMarMar09}.
Mass-loss rates are quite uncertain: for the O star the estimates range from $\dot{M}=8.7\times10^{-7}$\,M$_\odot$\,yr$^{-1}$ \citep{PitDou06} to $\dot{M}=8\times10^{-6}$\,M$_\odot$\,yr$^{-1}$ \citep{DouBeaCla05}; here we also consider $\dot{M}_\mathrm{O}=1\times10^{-6}$\,M$_\odot$\,yr$^{-1}$ as a representative value because most literature estimates tend toward the lower end of this range \citep[e.g.][]{PolCorSte21}.
For the WR star $\dot{M}_\mathrm{W} = 10^{-5.5}$\,M$_\odot$\,yr$^{-1}$ is a typical value \citep{WilMarMar09}.
Two cases are listed in Table~\ref{tab:wr140} corresponding to low and medium values of $\eta$, i.e, $\eta=0.022$ and $\eta=0.044$, respectively.
The low-$\eta$ parameters are the extreme values from the parameter study of \citet{PitDou06}, and the mid-$\eta$ values the preferred result from the X-ray analysis of \citet{SugMaeTsu15}.

\begin{figure}
\centering
  \includegraphics[width=\columnwidth]{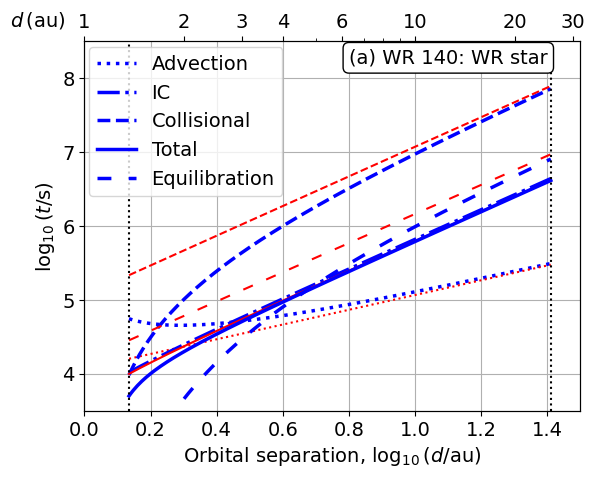}
  \includegraphics[width=\columnwidth]{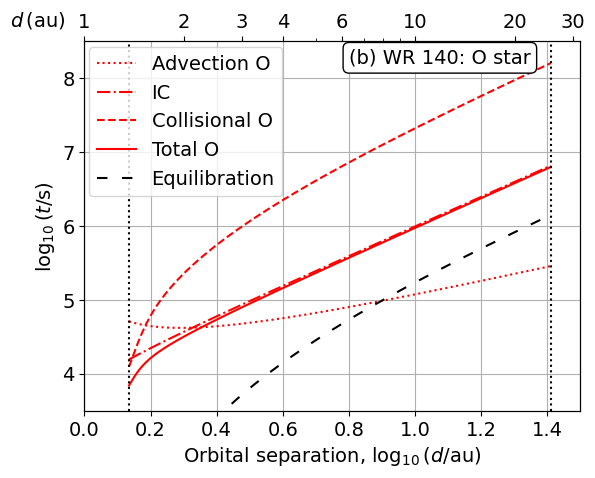}
  \caption{
      Timescales for cooling and advection in the shocked wind of the WR star (a) and O star (b) components of the binary system WR\,140, as a function of orbital separation, $d$, of the two stars.
    Vertical dotted lines show $d$ at periastron and apastron.
    The lower $x$-axis is $\log$-scaled whereas the upper $x$-axis shows the equivalent linear value of $d$.
    These plots assume the low-$\eta$ set of parameters (see text for details).
    In panel (a), the red curves assume the WR wind is not decelerated by the O star's radiation field, whereas the heavy blue curves assume that it is (more detail in the text).
  }
  \label{fig:tcool_lowe}
\end{figure}

\begin{figure}
\centering
  \includegraphics[width=\columnwidth]{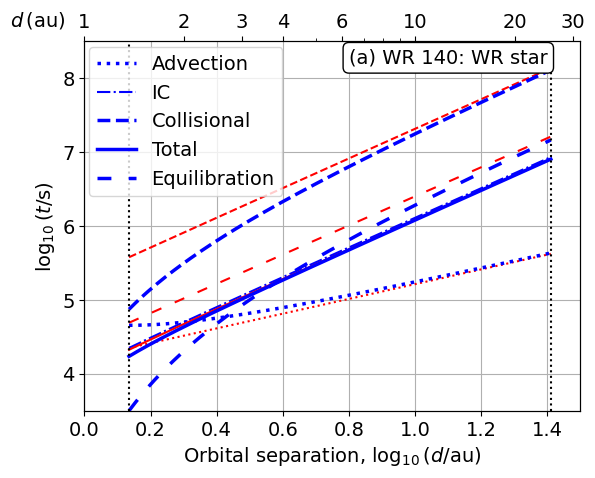}
  \includegraphics[width=\columnwidth]{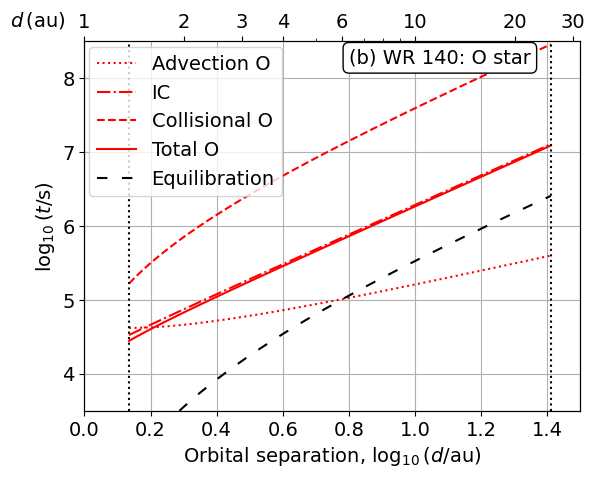}
  \caption{As Fig.~\ref{fig:tcool_lowe} but assuming the mid-$\eta$ set of parameters (see text for details).
  }
  \label{fig:tcool_mide}
\end{figure}

\subsection{Cooling time as function of orbital separation}
Here we consider the impact of the combined radiation fields of the O and WR stars on the shocked gas from each wind, as a function of the orbital separation, $d$, of the two stars.
We use $\eta$ to estimate the shock location and calculate the IC cooling associated with the radiation from both stars, as a function of $d$, as in the previous section.

The timescales are plotted in Fig.~\ref{fig:tcool_lowe} for the above-listed parameters using the low-$\eta$ set of mass-loss rates (for which $\eta=0.022$).
For both stars we use $\tau_\mathrm{adv}$ for the O star because the WR star has a much stronger wind and so the curvature radius of the shocked region is determined by the distance to the O star, not the WR star.
In panel (a) we assume with the red curves that the WR wind is not decelerated by the O star's radiation, whereas in the blue curves we assume that it is according to
\begin{equation}
V_\mathrm{W}(d_\mathrm{O}) = V_{\infty,\mathrm{W}} \left(1 - \frac{R_{\star,\mathrm{O}}}{d_\mathrm{O}} \right) \;,
\end{equation}
where $d_\mathrm{O}$ is the distance to the O star.
The latter case is what is included in the multi-dimensional simulations below, whereas reality is probably somewhere in between these two extremes.

For the shocked WR wind $\tau_\mathrm{ic}$ is $\sim30\times$ shorter than $\tau_\mathrm{ff}$, and the shock is radiative near periastron.
In the case where the WR wind is decelerated, $\tau_\mathrm{adv}$ is longer near periastron and the shock is radiative over a larger range of radii.
For the O star the shock should be radiative at periastron without IC cooling, but only at periastron.
With the inclusion of IC cooling the region of parameter space where the shocks are radiative is significantly increased to $d\approx2$\,au ($3\times10^{13}$\,cm).

Results for the mid-$\eta$ set of mass-loss rates (everything else kept constant) are shown in Fig.~\ref{fig:tcool_mide}.
Here we see that, because the wind collision region is moved further from the O star, the IC cooling timescale is about $3\times$ longer than in the low-$\eta$ case, and so the shocks should be radiative over a shorter range of orbital separations.

For both low-$\eta$ and mid-$\eta$ cases the IC cooling timescale is much shorter than the collisional-cooling timescale, $\tau_\mathrm{cie}$, and so one cannot obtain the correct physical result without considering IC cooling.
A key observation of the WR\,140 system is that the X-ray emission drops dramatically near periastron \citep{PolCorSte21} and line emission from intermediate ionization elements increases, indicative of a transition from an adiabatic to a radiative shock.
IC cooling can potentially explain both of these observations, in that energy is lost from the hot phase through IC cooling and does not emerge at X-ray energies (although there may still be significant X-ray emission from the partially-cooled and compressed gas), and the shock is expected to be radiative in the range of separations encountered near periastron.

\subsection{Post-shock equilibration of electrons and ions}
\label{sec:equil}
  Figs.~\ref{fig:tcool_lowe} and \ref{fig:tcool_mide} also plot the electron-ion equilibration time, $\tau_\mathrm{eq}$, for the postshock plasma of the two winds.
Looking first in Fig.~\ref{fig:tcool_lowe} at the low-$\eta$ wind parameters we see that $\tau_\mathrm{eq}$ for the shocked O-star wind is always significantly shorter than the cooling timescale.
On the other hand the timescale for the WR wind, $\tau_\mathrm{eq,W}$, is longer than both the cooling and advection timescales for the full orbit if the WR wind is not decelerated.
  In this case the IC cooling timescale is limited by the collisional heating rate of the electrons and will coincide with $\tau_\mathrm{eq,W}$.
IC cooling can cool the electrons faster than they are heated, and so they may never reach the equilbrium post-shock temperature.
In case the wind is effectively decelerated (the heavy blue lines) then $\tau_\mathrm{eq,W}$ is much shorter than $\tau_\mathrm{ic}$ near periastron, and for all parts of the orbit where $\tau_\mathrm{cool}<\tau_\mathrm{adv}$.
In this case, temperature equilibration is not limiting IC cooling efficiency in the parts of the orbit where radiative shocks are predicted.

  Similarly, Fig.~\ref{fig:tcool_mide} for the mid-$\eta$ wind parameters shows that $\tau_\mathrm{eq,O}\ll \tau_\mathrm{ic}$ for all separations in the shocked O-star wind, but $\tau_\mathrm{eq,W}>\tau_\mathrm{ic}$ far from periastron for both decelerated and free-flowing WR wind.
  In the parts of the orbit where radiative shocks are predicted, $\tau_\mathrm{cool}>\tau_\mathrm{eq,W}$ when the WR wind is decelerated, but not when the WR wind is free-flowing.
  For the assumptions made in the multi-dimensional simulations in Sections~\ref{sec:2dsim} and \ref{sec:wr140-3d} (efficiently decelerated WR wind) the electron-ion equilibration is not limiting the efficiency of IC cooling when radiative shocks are predicted.

  These results agree with the conclusion of \citet{PolCorSte05} that the collisional coupling of electrons and ions is weak in the shocks of WR\,140 over most of the orbit.
  The only difference is that we include effects of wind acceleration (and deceleration), and this modifies the timescales near periastron.
  We find that $\tau_\mathrm{eq,O}$ is by far the shortest timescale in the shocked O-star wind near periastron and a single-temperature fluid is a valid approximation.
  On the other hand, near apastron $\tau_\mathrm{eq,O}>\tau_\mathrm{adv}$ and temperature equilibration will not occur in this part of the orbit.
  For the WR star, we assume its wind is decelerated as it approaches the O star near periastron and, for this assumption, $\tau_\mathrm{eq,W}$ is the shortest timescale and efficient electron-ion temperature equilibration occurs.
  Far from periastron the collisional coupling is weak and temperature equilibration does not occur.

\subsection{2D simulation of wind-wind collision in WR 140}
\label{sec:2dsim}
\begin{figure}
\centering
  \includegraphics[width=0.85\columnwidth]{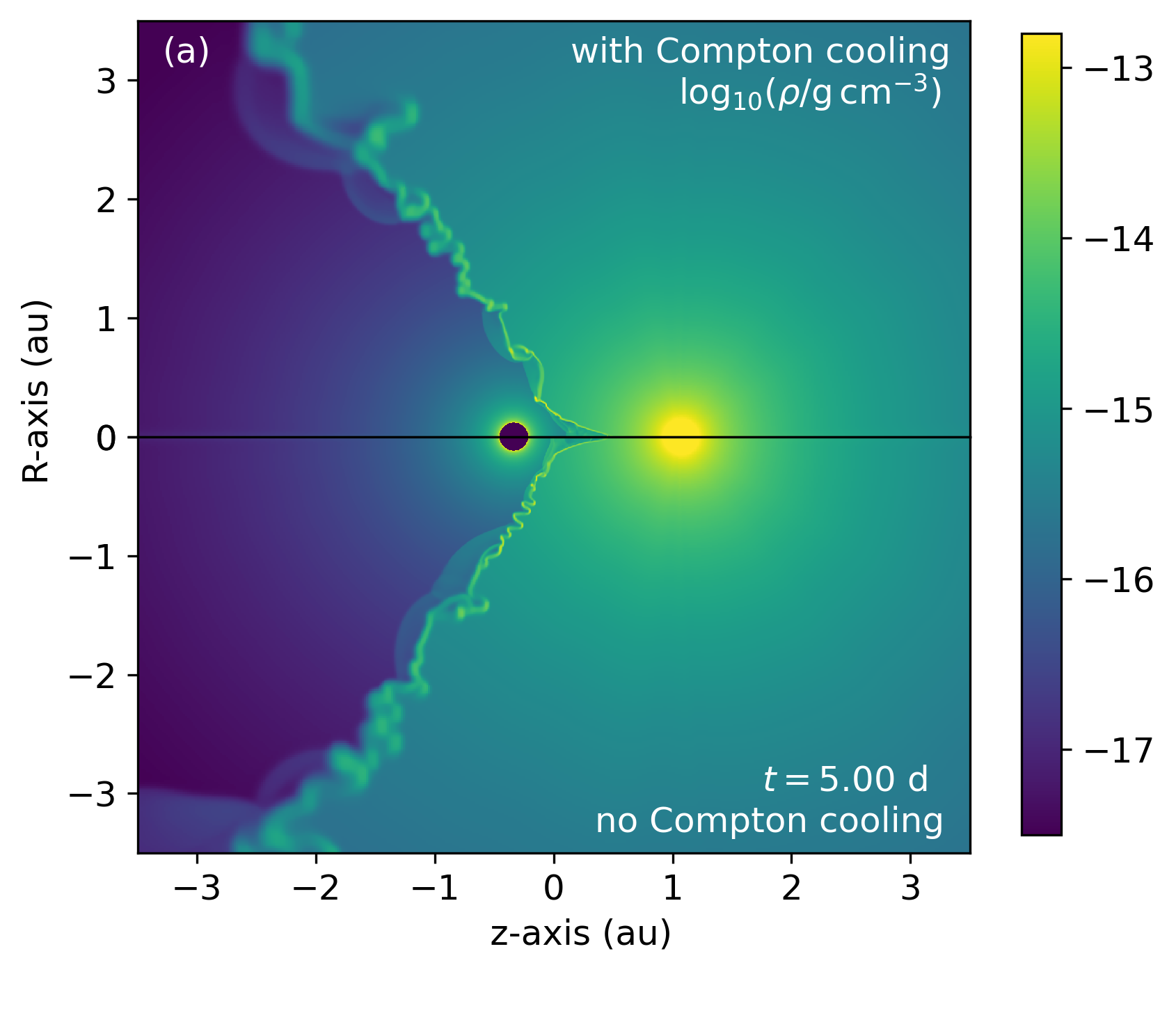}
  \includegraphics[width=0.85\columnwidth]{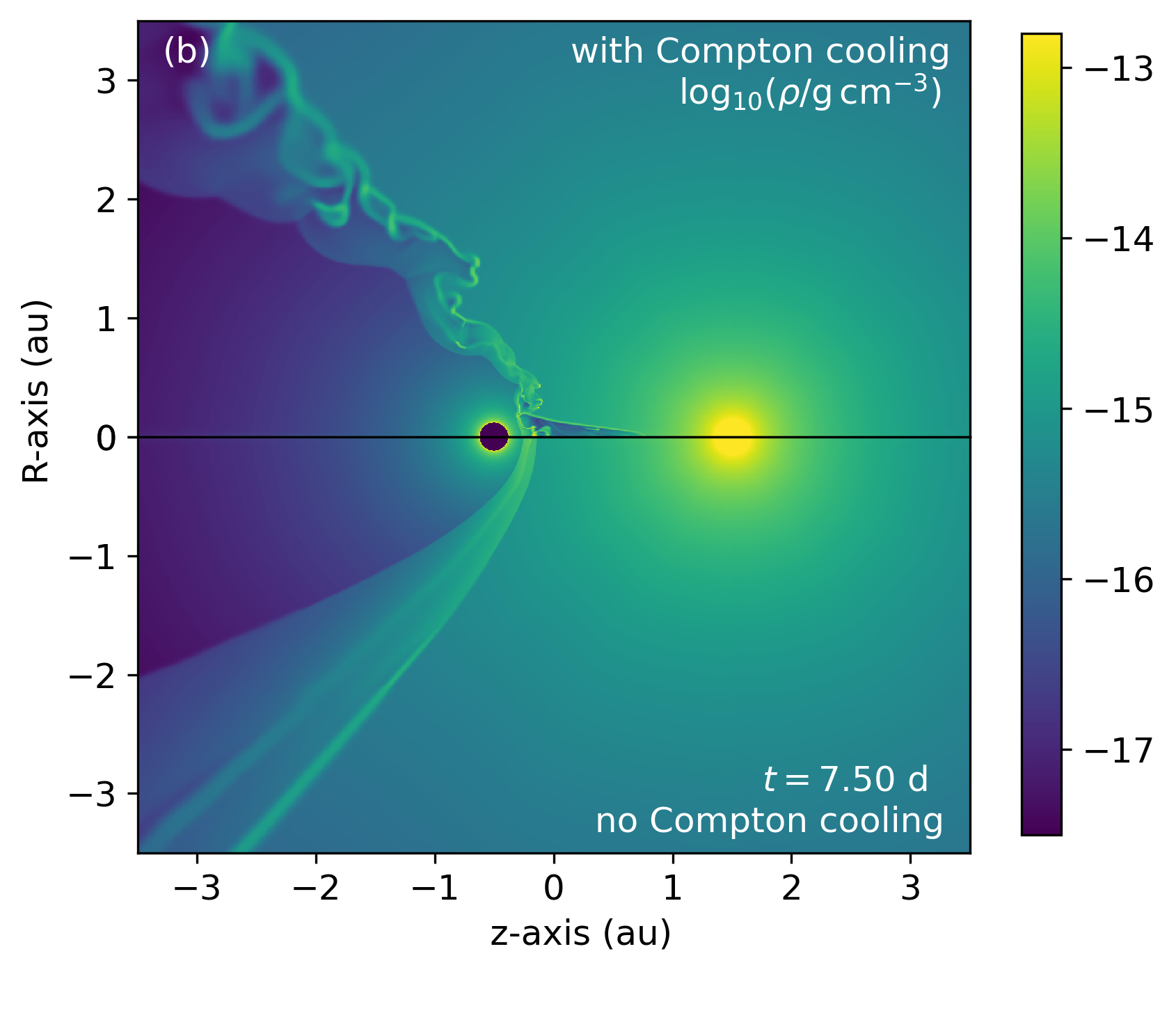}
  \includegraphics[width=0.85\columnwidth]{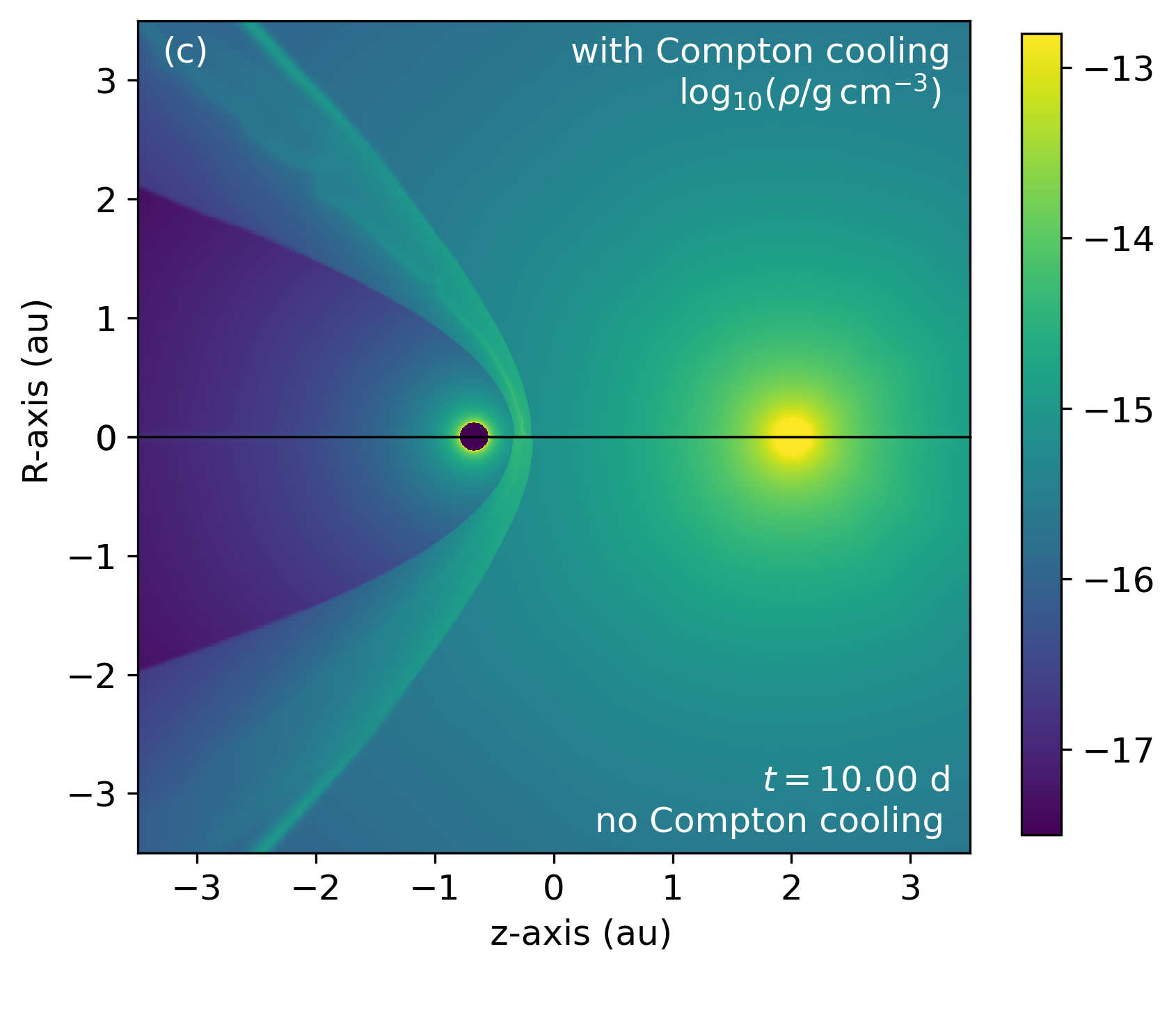}
  \caption{Gas density for 2D simulations of wind-wind collision for the WR\,140 system using low-$\eta$ mass-loss rates, with IC cooling included (upper half-plane) and omitted (lower half-plane) for stellar separation (a) $d=2.1\times10^{13}$\,cm (1.4\,au), (b) $d=3\times10^{13}$\,cm (2.0\,au) and (c) $d=4\times10^{13}$\,cm (2.7\,au).
  }
  \label{fig:wr140-2d-lowe}
\end{figure}

\begin{figure}
\centering
  \includegraphics[width=0.85\columnwidth]{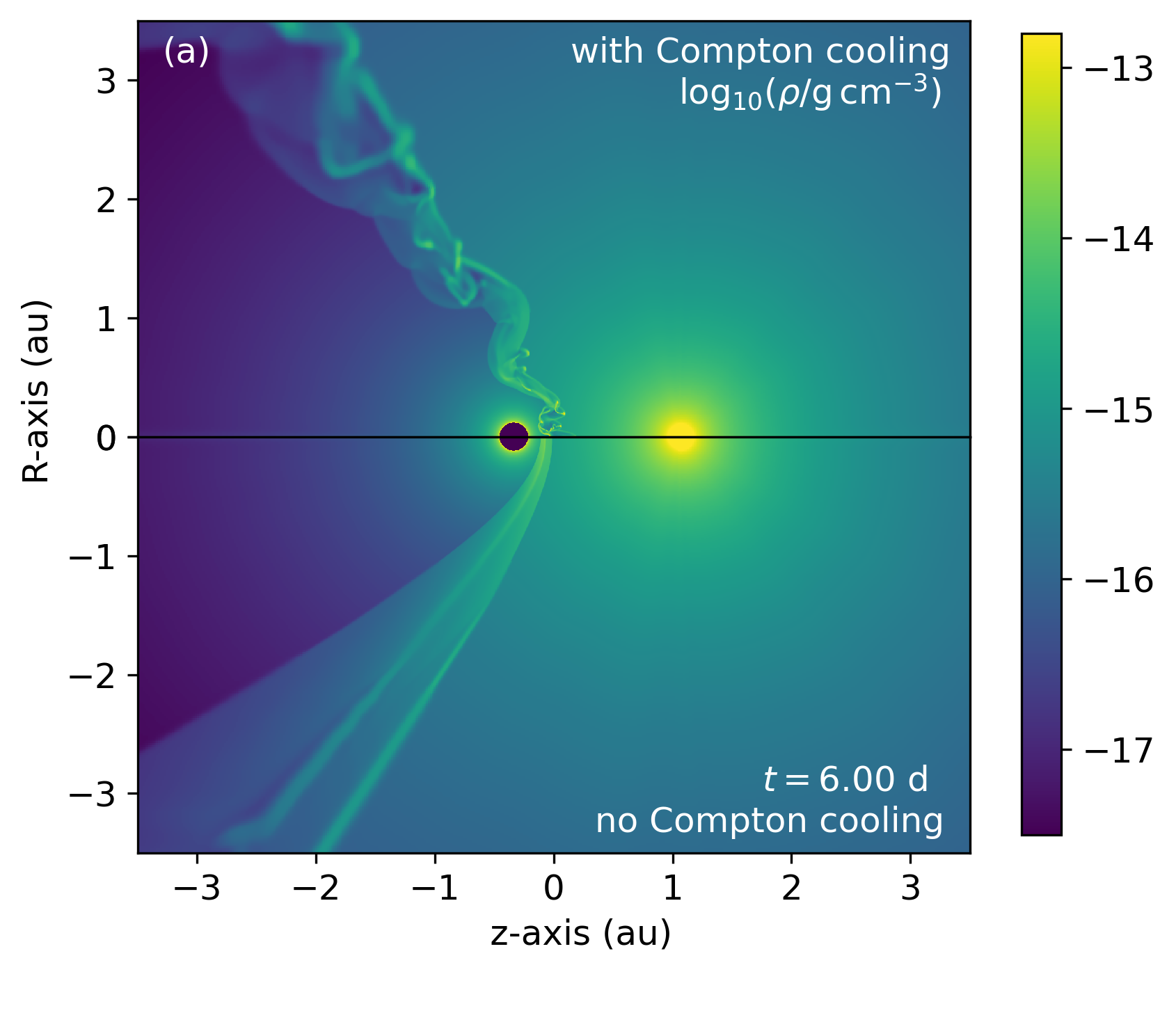}
  \includegraphics[width=0.85\columnwidth]{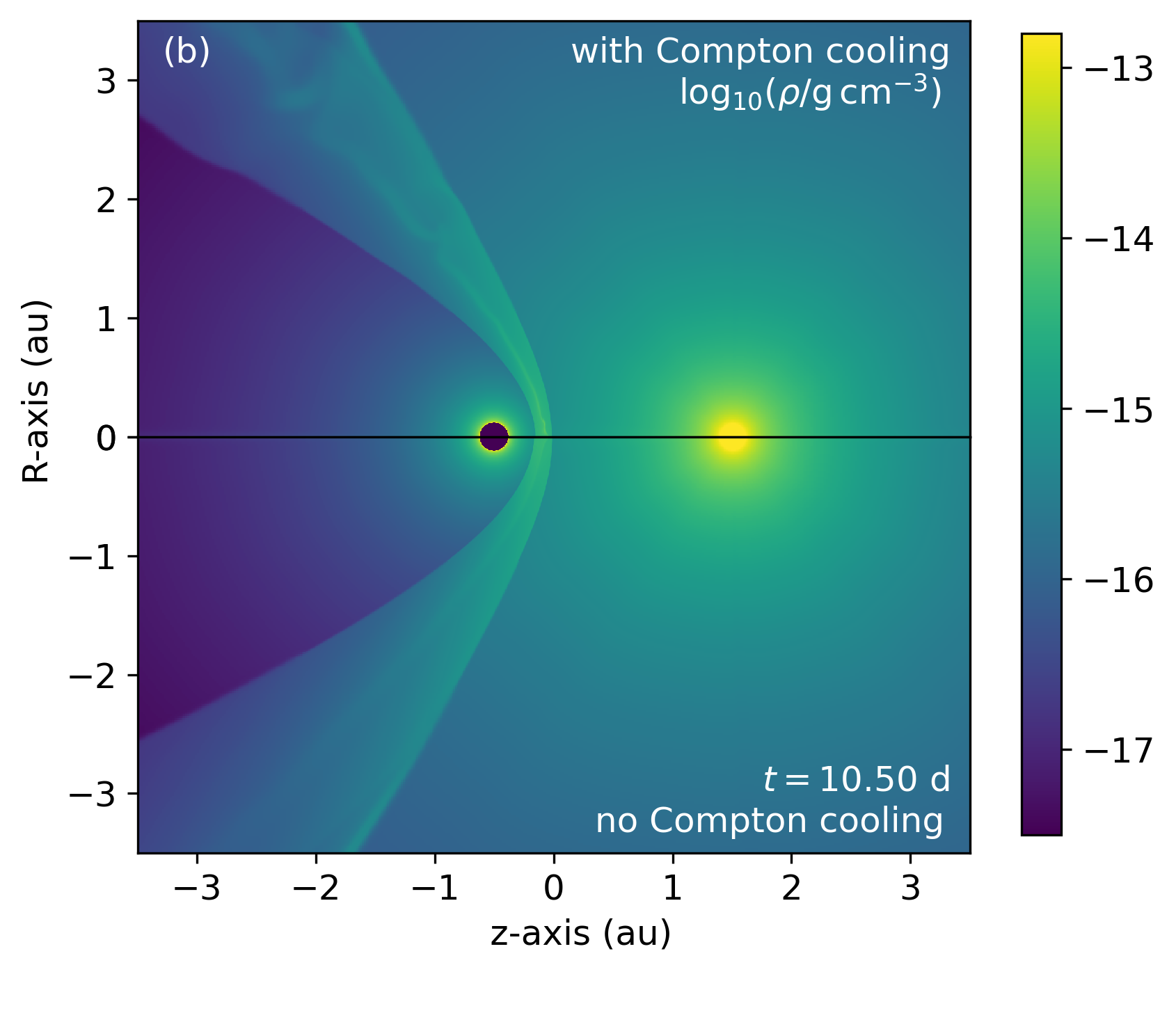}
  \caption{Gas density for 2D simulations of wind-wind collision for the WR\,140 system using mid-$\eta$ mass-loss rates, with IC cooling included (upper half-plane) and omitted (lower half-plane) for stellar separation (a) $d=2.1\times10^{13}$\,cm (1.4\,au) and (b) $d=3\times10^{13}$\,cm (2.0\,au).
  }
  \label{fig:wr140-2d-mide}
\end{figure}

We used the (magneto-)hydrodynamics code \textsc{pion} \citep{MacGreMou21} to make 2D simulations of the wind-wind collision of the two stars in the WR\,140 system for different stellar separations, using the 2D setup described in section~\ref{sec:sim-methods}.
The cylindrical $R$-$z$ plane is simulated (with rotational symmetry in the angular direction), and the two stars are static and on the $z$-axis.
Simulations were run for separations ranging from $d=1.4$\,au ($2.1\times10^{13}$\,cm; approximately periastron for WR\,140) to $d=6.7$\,au ($1\times10^{14}$\,cm).

The outer (coarse) grid has $z\in [-5,5] \times10^{14}$\,cm, $R\in [0,5]\times10^{14}$\,cm and $512\times256$ cells, and a further 5 nested grids centred on the origin.
Each of these is identical to the coarse grid except successively with cell diameter $\Delta x$ a factor of 2 smaller than the level above.
The finest grid has $z\in [-1.5625,1.5625]\times10^{13}$\,cm (or $\pm1.04$\,au), $R\in [0,1.5625]\times10^{13}$\,cm and $\Delta x\approx6.1\times10^{10}$\,cm (0.004\,au).
The O-star is placed at $z=-d/3$ and the WR star at $z=2d/3$, and the stars are assumed to be unmagnetized and to emit a spherically symmetric wind.
The simulations are run with a CFL parameter of 0.3, and snapshots are saved every 0.25\,d until the simulations reaches a stationary state or fully-developed unstable flow.
Simulations were run for the low-$\eta$ and mid-$\eta$ set of mass-loss rates (see Table~\ref{tab:wr140}).

Fig.~\ref{fig:wr140-2d-lowe} shows results for the low-$\eta$ mass-loss rates, comparing runs with IC cooling (upper half-plane) and without (lower half-plane) for separations $d=[2.1,3,4]\times10^{13}$\,cm  (1.4, 2.0 and 2.7\,au, respectively) from top to bottom.
Both simulations are strongly unstable in panel (a) corresponding to periastron, signifying radiative shocks at the apex of the wind-wind collision.
This is expected from Fig.~\ref{fig:tcool_lowe}, which shows that the O-star shocked wind should be radiative with and without IC cooling at periastron.
For larger separations only the run with IC cooling is unstable, again as expected from Fig.~\ref{fig:tcool_lowe}.
For $d=4\times10^{13}$\,cm (panel c) neither simulation is unstable; the run with IC cooling shows an overdense layer near the contact discontinuity indicating that some cooling has taken place, but not sufficient for runaway cooling to set in.
For the run without IC cooling this layer is much less prominent, indicating that both shocks are effectively adiabatic, as expected from Fig.~\ref{fig:tcool_lowe}.

The unstable simulations in Fig.~\ref{fig:wr140-2d-lowe} show prominent spikes along the symmetry axis.
These arise because of the imposed rotational symmetry and the coordinate singularity: no gas can cross the cell boundary $R=0$ because its area is zero.
Once a dense cooled clump forms on the symmetry axis there is no force that can move it off the axis and so it just moves along the line $R=0$.
The large inertia of this clump means that its position has a large amplitude of oscillation, seeding the formation of the spikes \citep[see also, e.g.,][]{GreMacHaw19}.

Fig.~\ref{fig:wr140-2d-mide} shows results for the mid-$\eta$ mass-loss rates, this time only for $d=2.1\times10^{13}$\,cm (panel a) and $d=3\times10^{13}$\,cm (panel b) (1.4 and 2.0\,au, respectively).
As expected from Fig.~\ref{fig:tcool_mide}, the runs without IC cooling do not have radiative shocks near the apex of the shock and so remain dynamically stable.
By contrast, at periastron the shocked layer is very unstable for the run with IC cooling.
The range of separations where $\tau_\textrm{cool}<\tau_\mathrm{adv}$ is smaller for the mid-$\eta$ case than for the low-$\eta$ case, and so already at $d=3\times10^{13}$\,cm the run with IC cooling is also non-radiative, albeit with some weak instability and an overdense layer at the contact discontinuity.
This agrees with panel (a) of Fig.~\ref{fig:tcool_mide} that shows  $\tau_\textrm{cool} = \tau_\mathrm{adv}$ at $d\approx3\times10^{13}$\,cm for the WR star wind.

Overall, these simulations agree very well with the expectations from simple analytic estimates of the relevant timescales.
IC cooling is required in order to obtain the correct thermal behaviour of the post-shock gas for the wind-collision region of WR\,140 near periastron.

\begin{figure*}
\centering
  \includegraphics[width=\textwidth]{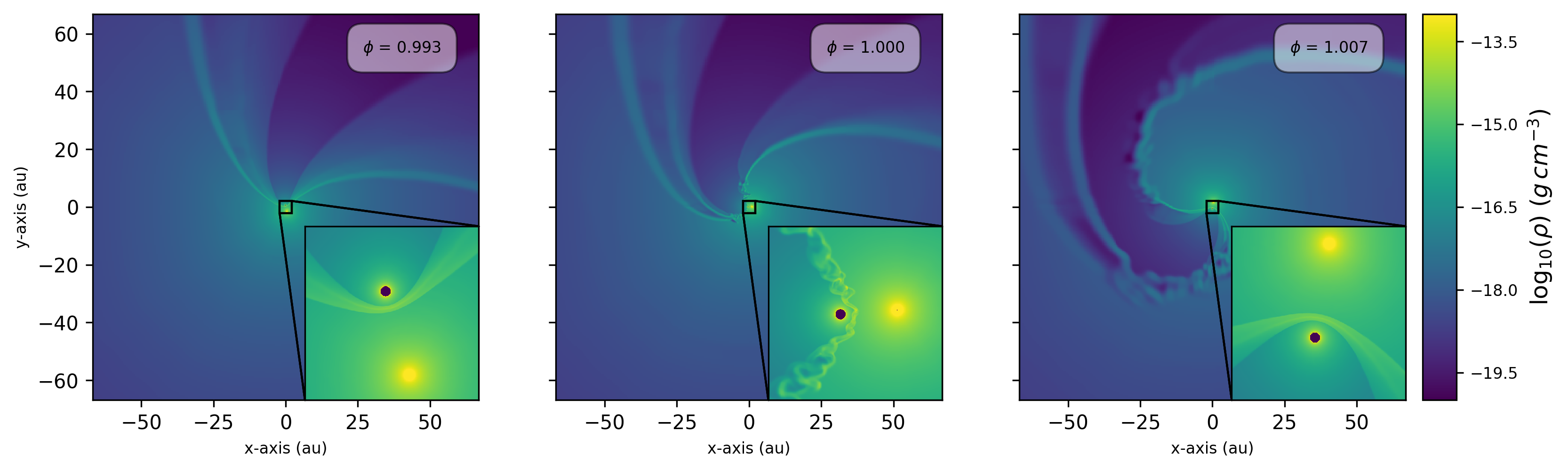}
  \caption{Gas density on a logarithmic colour scale in a slice through the orbital plane $z=0$ for the WR\,140 system using mid-$\eta$ mass-loss rates, with IC cooling included (simulation \texttt{mhd-2}).
  The left-hand panel shows a time shortly before periastron (orbital phase $\phi=0.993$), the middle panel almost exactly at periastron, and the right panel just after periastron ($\phi=1.007$).
  The inset shows a zoomed-in view of the wind-collision region.
  }
  \label{fig:wr140-3d-mide-comp}
\end{figure*}

\begin{figure}
\centering
  \includegraphics[width=0.9\columnwidth]{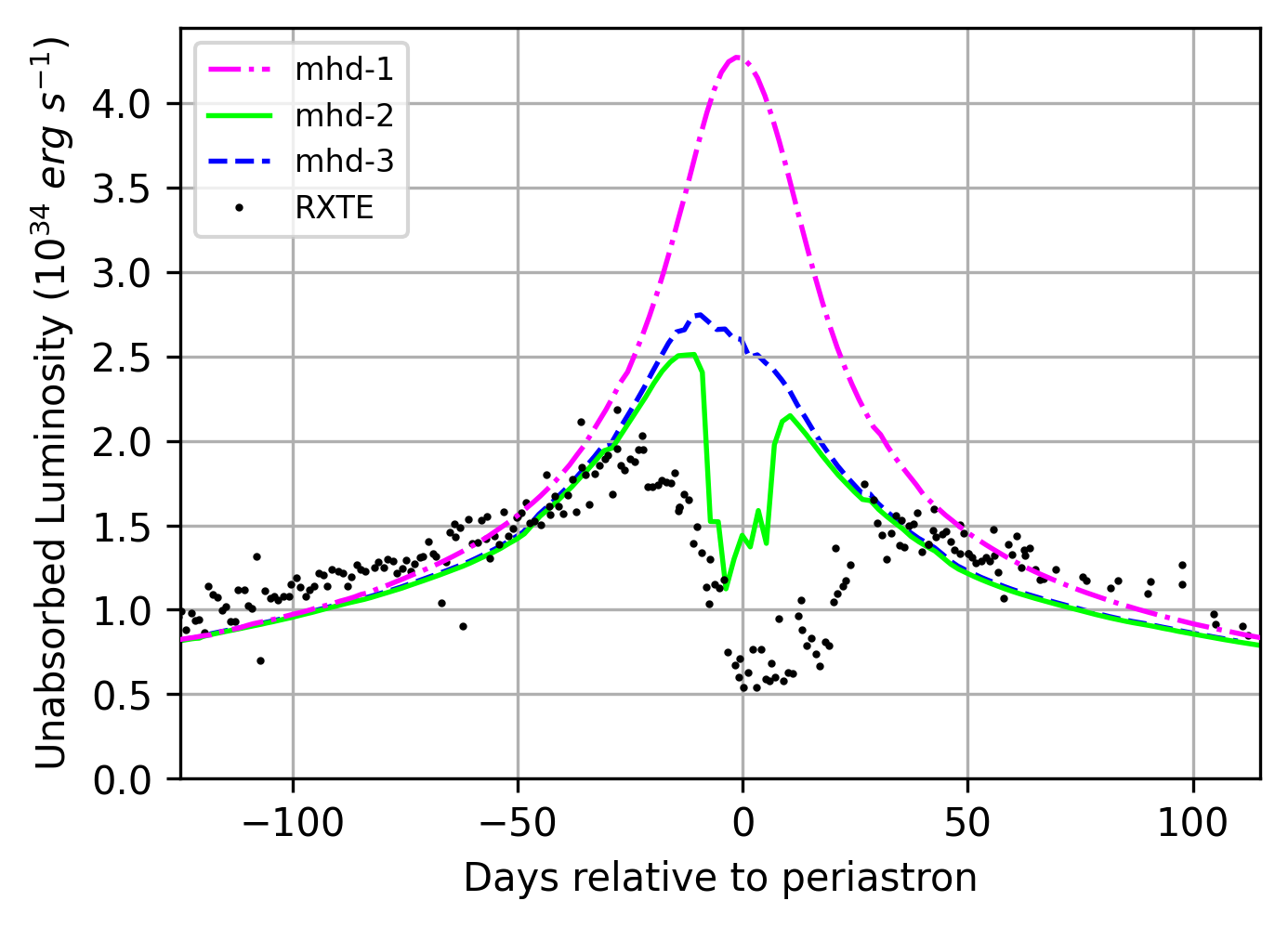}
  \caption{X-ray luminosity (2-10\,keV) of the 3D simulation as a function of time near periastron for the simulations without (\texttt{mhd-1}, \texttt{mhd-3}) and with (\texttt{mhd-2}) IC cooling, and without (\texttt{mhd-1}) and with (\texttt{mhd-2}, \texttt{mhd-3}) wind acceleration.
  The RXTE phase-folded and absorption-corrected hard-X-ray light-curve is plotted as the black points \citep[from][]{PolCorSte21}.
  }
  \label{fig:wr140-3d-mide-xray}
\end{figure}

\subsection{3D simulation of wind-wind collision in WR 140}
\label{sec:wr140-3d}

To obtain more realistic results we also ran 3D MHD simulations of the WR\,140 system around periastron using \textsc{pion}, similar to calculations by \citet{EatPitVan22a,EatPitVan22b} but again including IC cooling and not dust cooling.
A static nested grid was used with 7 levels of refinement and $256\times256\times64$ grid cells on each level.
The stellar orbits were solved using a leapfrog integrator with initial conditions (position and velocity of the two stars) pre-calculated and set as parameters.
The coarsest grid has a domain $[x,y] \in \pm10^{15}$\,cm ($\pm67$\,au) and $z \in \pm2.5\times10^{14}$\,cm ($\pm16.7$\,au) and the finest grid $2^6\times$ smaller in each dimension: $[x,y] \in \pm1.5625\times10^{13}$\,cm ($\pm1.04$\,au) and $z \in \pm3.90625\times10^{12}$\,cm ($\pm0.26$\,au), with $\Delta x = 1.22\times10^{11}$\,cm (0.0082\,au).
A CFL parameter of 0.1 was used for the dimensionally unsplit integration scheme.
We ran simulations both with and without IC cooling, and also at a factor of 2 lower resolution to assess convergence.

The simulations were run with the mid-$\eta$ set of parameters.  For the stellar magnetic field, we follow \citet{MacGreMou21} by imposing a split-monopole field emerging from the stellar surface, with a surface field of 1\,G for the O star and 100\,G For the WR star, such that the winds have Alfv\'en Mach numbers $\approx50$ and $\approx350$ at the wind injection boundary, respectively.
Simulation \texttt{mhd-1} does not include IC cooling or wind acceleration (i.e., winds are injected at the terminal velocity) and so is a control run to assess the effects of IC cooling.
Simulation \texttt{mhd-2} includes IC cooling and wind acceleration.
Simulation \texttt{mhd-3} includes wind acceleration but not IC cooling.
The lower-resolution runs are discussed in Appendix~\ref{app:resolution}.
Simulation \texttt{mhd-1} was started at orbital phase 0.89, and simulations \texttt{mhd-2} and \texttt{mhd-3} at phase 0.95 (about 150 days before periastron).
This is long enough that any imprint of the initial conditions has been swept far from the inner part of the orbit well before periastron.

Snapshots showing slices through the midplane $z=0$ are shown in Fig.~\ref{fig:wr140-3d-mide-comp} for the mid-$\eta$ mass-loss rates just before (a), during (b) and after (c) periastron.
The shocks become radiative shortly before periastron, visible from the high-density sheets at the contact dicontinuity and termination shock of the WR wind.
These are, respectively, the cooled wind from the O star and the WR star.
We do not see much instability in the flow before periastron, in contrast to the results of \citet{EatPitVan22b}, probably because we do not include the dust cooling that is effective in these sheets.
IC cooling is only effective very close to the stars, whereas the dust cooling is a density-dependent process.
Magnetic fields may also play a role in stabilising the flow, because a magnetised fluid is much less susceptible to Kelvin-Helmholtz instability \citep{FraJonRyu96}.

At periastron the shocks in the wind-collision region become radiative and dynamically unstable, seen in the inset to the middle panel of Fig.~\ref{fig:wr140-3d-mide-comp}.
The shocked region collapses to a thin sheet that develops non-linear corrugations that are then swept away from the stagnation point.
These are also seen in the large-scale flow in the right-hand panel, where the shocked gas that is swept away from the binary system appears turbulent and disturbed.
The wind-collision region remains radiative for only a short time around periastron and then recovers its quasi-adiabatic behaviour and dynamical stability (inset of right-hand panel of Fig.~\ref{fig:wr140-3d-mide-comp}).

Fig.~\ref{fig:wr140-3d-mide-xray} shows the X-ray luminosity of the mid-$\eta$ 3D simulations as a function of time with respect to periastron.
The luminosity in the 2-10\,keV band is calculated using the method of \citet{GreMacKav22} for models of bow shocks, using emissivity tables generated with \textsc{xspec} v12.9.1 \citep{Arn96}.
The first $\approx20$\,d of the simulated light-curve is omitted because the initial conditions are still affecting the X-ray emission.
The control simulation \texttt{mhd-1} (without IC cooling or wind acceleration) follows the expected quasi-adiabatic behaviour ($L_\mathrm{X}\propto d^{-1}$) because the shocks never become strongly radiative.
The simulation with IC cooling (\texttt{mhd-2}) shows a slower increase in luminosity compared with the control simulation (\texttt{mhd-1}) from 100 to 20 days before periastron.
The same result is seen in simulation \texttt{mhd-3}, demonstrating that this is due to wind acceleration: as the shocks move closer to the O star the shock speed and post-shock temperature decrease, reducing the hard-X-ray emission.
Simulation \texttt{mhd-3} departs from the adiabatic limit but shows a smooth peak before periastron followed by a steady decrease, returning to the same luminosity as \texttt{mhd-1} around 100 days after periastron.
Simulation \texttt{mhd-2} shows a sharp drop in X-ray luminosity near periastron, as the additional IC cooling makes the shocks become radiative, and the emission recovers again some days after periastron.
This dip is again asymmetric with respect to periastron, as found by previous authors; the post-shock pressure is larger before periastron because the stars are approaching each other and so the flow speed through the shock is larger than after periastron \citep{PitPar10}.

The absorption-corrected data from RXTE from \citet[][fig.~6]{PolCorSte21} are overplotted for reference on Fig.~\ref{fig:wr140-3d-mide-xray}; these are phase-folded observations for the periastron passages of 2001 and 2009.
The simulation with IC cooling provides a much better qualitative match to the observations than without, although we have not tuned the simulation parameters at all to obtain this match (apart from choosing wind parameters recommended by \citealt{SugMaeTsu15}).
Far from periastron the simulated X-ray emission is about 20 per cent too low, indicating that the wind density is slightly too low in the wind-collision region.
To match the data while maintaining the same value of $\eta$ we would need to increase the mass-loss rates of both stars by a small amount (about 10 per cent).
This would have the effect of triggering radiative shocks slightly earlier before periastron, and the shocks would remain radiative for longer (agreeing better with the observational data).

It is encouraging that this first attempt to model the X-ray emission is so close to the observations and, in particular, the asymmetry and shape of the dip in emission matches the observations very well.
We note that any simulation producing radiative shocks at the stagnation point will likely have a similar asymmetric dip in the hard X-ray lightcurve, \citep[e.g.][]{PitPar10, RusCorOka11}, but the the 3D simulations presented here show that IC cooling should be included in order to correctly determine whether the shocks become radiative or not, the time of onset and the duration of the dip.

\section{Discussion}
\label{sec:discussion}
  A necessary, though not sufficient requirement for dust formation in CWBs, is that the shocked plasma cools from the postshock temperature, $T\sim10^7-10^8$\,K to lower temperatures.
  Cooling proceeds approximately isobarically, and so the collisional-cooling timescale decreases as $T$ decreases and $\rho$ increases.
  Once collisional cooling becomes efficient, a runaway process down to $T\sim10^4$\,K develops, with the bottleneck at the highest temperatures.

  Given the best observational estimates of the wind properties in WR\,140, we showed that the postshock plasma should not be able to cool when only collisional processes are included, and that (near periastron) IC cooling can increase the cooling rate sufficiently to drive the shocks into the radiative regime, allowing cooling to $\sim10^4$\,K.
  Further cooling to $\sim10^3$\,K requires the gas to become self-shielding, allowing molecule and eventually dust formation, and our simulations have neither the dynamic range nor the chemical kinetics needed to model this.
  Nevertheless we argue that IC cooling is a necessary ingredient in any shock-cooling model for WR\,140 (and other CWBs) that aims to to determine when and where dust is produced during the orbit.

While WR\,140 provides an excellent case study of the importance of IC cooling, there are many other WR binary systems, of which some produce dust and some do not.
For example, WR\,104 is a persistent dust producer from a closer binary in a more circular orbit (period 200\,d) with separation $d\approx3\times 10^{13}$\,cm or 2\,au \citep{TutMonDan99}, which has orbital, wind and stellar parameters that put it in the IC cooling regime.
The long-period system WR\,112, in contrast, must be in the regime of line-driven radiative cooling because of the relatively slow wind and large mass-loss rate of the WC star, together with the relatively large wind momentum ratio, $\eta\approx0.13$ \citep{LauHanHan20}.

\begin{figure}
\centering
  \includegraphics[width=\columnwidth]{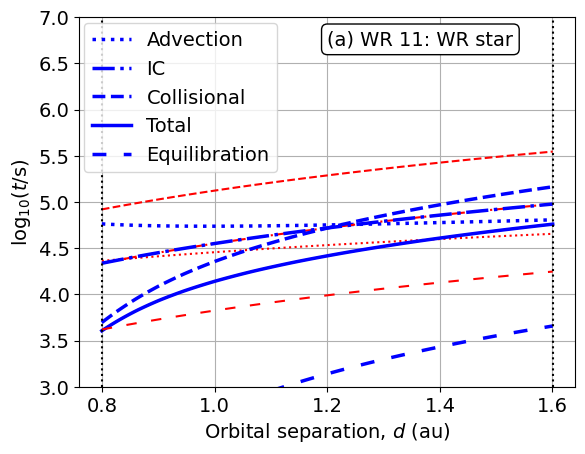}
  \includegraphics[width=\columnwidth]{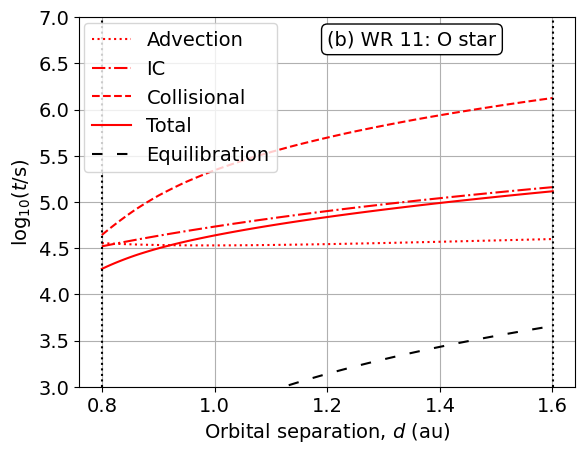}
  \caption{Timescales for cooling and advection in the shocked wind of the WR star (a) and O star (b) components of the binary system WR\,11 ($\gamma^2$ Vel), as a function of orbital separation, $d$, of the two stars.
    For panel (a) the heavy blue lines assume the WR wind is decelerated by the O star's radiation field, whereas the light red lines assume no deceleration.
    Vertical dotted lines show $d$ at periastron and apastron.
  }
  \label{fig:tcool_wr11}
\end{figure}

The nearby CWB $\gamma^2$ Velorum (WR\,11) has a wind collision zone with similar orbital separation as WR\,140 at periastron, but does not produce any dust \citep[see e.g., the discussion in][]{LauEldHan20}.
The timescales in the wind collision region of this system are plotted in Fig.~\ref{fig:tcool_wr11}, using the wind parameters from \citet{LamMilLie17} and stellar parameters from \citet{ReiKisRei17}.
We see that IC cooling dominates over collisional processes, and should induce sufficient cooling so that radiative shocks are present for at least part of the orbit.
Why $\gamma^2$ Velorum does not then produce dust is not clear, although dust formation in such extreme environments is not well-understood \citep{Che15}, being dependent on the radiation field, gas density, composition, and dynamical timescale of the gas.
Realistic modelling of the physical conditions in the wind-collision zone is a required input to any theoretical model for dust formation, and Fig.~\ref{fig:tcool_wr11} shows that IC cooling is an important ingredient.

The WR binaries discovered in the Small \citep{FoeMofGue03a} and Large \citep{FoeMofGue03b} Magellanic Clouds are predominantly short-period systems with $P\sim2-30$\,d, in contrast to the Galactic WR binaries (although there are a few long period binaries with $P\sim100$\,d in the Large Magellanic cloud; \citealt{SheSabHai19}).
\citet{SheHaiTod16} showed that the WR component in the binaries are also systematically more luminous than their Galactic counterparts.
This, along with the weaker winds of low-metallicity WR stars \citep{SanVinHam20}, means that we should expect IC cooling to be even more dominant over collisional cooling processes than for the Galactic systems explored in this work.
A recent X-ray survey of the Tarantula Nebula massive stars \citep{CroBroTow22} provides an excellent dataset for comparison with models.

  Furthermore, CWBs consisting of two early O stars should also provide a good test of the importance of IC cooling, since they have weaker winds and may also be more luminous than the classical WR stars.
  This means that (all else being equal) the collisional cooling timescale is longer and the IC cooling timescale shorter than in systems like WR\,140.
  We plan to investigate this further in future work.

\subsection{Observational signatures}
An observational signature that IC cooling is the dominant coolant in a region is a reduction in the observed X-ray emission compared with expectations (especially hard X-ray emission from the hottest gas).
This process takes energy from the thermal electrons and adds it to the local radiation field, but the photon energy is only significantly changed if the electrons are relativistic (multiplied by $\gamma^2$, where $\gamma$ is the Lorentz factor of the electrons), or where a system is in the $y_\mathrm{c}\gg1$ multiple-scattering limit.
Such a variation can be detected as the Sunyaev-Zeldovich effect in galaxy clusters because the CMBR is so well characterised that even a tiny deviation from a blackbody can be measured with high significance.
It is very difficult to measure in the environment of massive stars, however, because the plasma is so dynamic and spectral lines are already significantly broadened and shifted by stellar winds and orbital motion, respectively.
The overall increase in energy density of the photon field is negligible compared with the stellar luminosity because the wind mechanical luminosity (much of which is converted to thermal energy in the shocks at the wind-collision zone) is generally only a small fraction of the radiative luminosity.
X-ray light curves are probably the only unambiguous measurement of the impact of IC cooling.

  A complicating factor is that in principle any cooling process may induce runaway radiative cooling, if its cooling timescale is shorter than the advection time, and this will also reduce the hard X-ray emission in the same way.
  For example wind clumping could locally reduce the collisional cooling timescale in overdense regions with respect to expectations for a smooth wind.
  A careful analysis would be required to exclude the possibility of efficient collisional cooling and definitively conclude that IC cooling is dominant.
  It is not clear whether the wind properties of WR\,140 can be constrained well enough to allow such a conclusion.
  This paper makes an initial exploration of the applicability of IC cooling to WR\,140 and potentially other CWB systems, showing that it should be considered and may in some cases be crucially important.

\subsection{Non-thermal radiation}
CWBs have been observed to emit non-thermal radiation at radio and $\gamma$-ray energies.
For example, synchrotron radio emission from WR\,140 was measured by \citet{DouBeaCla05} as a function of orbital phase of the system.
The extreme CWB system $\eta$ Car has been observed in $\gamma$-rays at GeV  \citep{TavSabPia09,FERMI_2010_EtaCar} and TeV \citep{HESS20_EtaCar} energies, as well as non-thermal hard X-rays \citep{SekTsuKit09, 2018NatAs...2..731H}.
Variation of the non-thermal X-ray \citep{2018NatAs...2..731H} and $\gamma$-ray \citep{WhiBreKon20, MarRei21} emission with orbital phase has been clearly detected.
Possible orbital variability at GeV energies has also been reported from the CWB $\gamma^2$ Velorum (WR 11) using \textit{FERMI} data \citep{MarReiLi20},

  The non-thermal radio emission is synchrotron radiation from relativistic electrons that depends on the local magnetic field strength, while the non-thermal X-rays are most plausibly produced by IC scattering off the same electrons. Diffusive shock acceleration is the presumed origin of the non-thermal emitting particles.
  Shock acceleration also depends on the magnetic field strength and orientation at the shock \citep[e.g.][]{Dru83,Bel11,WhiBreKon20}.
  Predicting acceleration and non-thermal radiation from CWBs therefore requires either assumptions about the field strength and configuration or simulation of the magnetic fields with the hydrodynamics.
  We have demonstrated a robust method to include magnetic fields in 3D simulations of CWBs, and in future work we will use the results from these and similar simulations to predict broadband non-thermal emission for quantitative comparison with observations such as the above results.

  These detections of non-thermal emission from CWBs raise the prospect that the next generation of high-energy observatories, especially the Cherenkov Telescope Array \citep{ActAgnAha11}, may be able to observe time-dependent non-thermal emisison from many CWB systems in the Galaxy.
  Detailed modelling of the phase-dependent properties of the shocks, the thermal plasma and magnetic fields, will be an important step towards understanding particle acceleration in these systems.

\subsection{Mass-loss rates of stars in CWBs}

For eccentric systems where IC cooling is the dominant process determining the transition from adiabatic to radiative shocks, this introduces an extra constraint on the properties of the system that could be used to better determine the mass-loss rates of the two stars.
The IC cooling time is independent of density (Equation~\ref{eqn:tau-comp}), depending only on the radiation energy density from the two stars.
This means that, to the extent that the orbital parameters, stellar luminosity and wind velocities are known, the location of the wind-collision region is determined at the orbital phase where the shocks switch from adiabatic to radiative, and so $\eta$ is determined without any knowledge of the wind density.
The X-ray emission far from periastron (in the adiabatic regime, where cooling is mainly from free-free emission) can also be used to measure the density in the wind-collision zone.
Because we know $\eta$, the wind density and the respective wind speeds, this determines the mass-loss rates of both stars.
Alternatively, if $\dot{M}$ for one of the stars can be measured accurately from optical/UV spectral analysis, then knowledge of $\eta$ determines $\dot{M}$ of the weaker component.
As far as we are aware this has not been proposed in the literature, and could provide an independent constraint on the otherwise degenerate estimates for the rate of mass-loss by clumped stellar winds.

The IC cooling timescale is independent of gas density, and so both the overdense clumps and the underdense surroundings will be equally cooled by this process.
With collisional cooling, the denser clumps cool more rapidly and may effectively decouple from the lower-density plasma which remains hot.
If IC cooling dominates to the extent that our results suggest, then wind clumping may not have significant consequences in the wind-collision zone.
All of the shocked plasma cools on the same timescale and will be compressed to the point that collisional cooling takes over.
The clumping in WR and O-star winds may nevertheless have some effect on our results: it is thought that the inertia of overdense clumps may allow them to penetrate through the wind-collision region, closer to the other star, increasing the gas pressure and potentially producing the X-ray flares observed in $\eta$ Car and (to a lesser extent) in WR binaries \citep{MofCor09}.
This broadens the wind-collision region, changing $U_\gamma$ in the shocked gas and therefore changing the IC cooling time.
The degree of clumping as a function of distance from the stellar surface is important here, whether driven by subsurface convection with weak clumping \citep{GraCheSan16, MoePonHen22} or the strong clumping inferred for line-driven winds \citep{BraDeKBes22, DriSunDag22}.

\section{Conclusions}
\label{sec:conclusions}
We have made a detailed investigation of the importance of IC cooling of the shocked thermal plasma in colliding-wind binary systems.
This process dominates over free-free cooling for a wide range of parameter space, and can be the determining factor as to whether or not a shock will be adiabatic or radiative.
Close binaries (or eccentric binaries near periastron) are expected to have radiative shocks and the parameter space with $\eta\ll1$ is dominated by IC cooling.
Estimates of whether a shock is radiative can be wrong both qualitatively and quantitatively if IC cooling is not included, especially for $\eta\ll1$ where the cooling time can be shortened by $10-100\times$ when IC cooling is considered.
The standard criterion of \citet{SteBloPol92} for determining whether a shock is radiative should be augmented with a further criterion based on IC cooling for close binaries with luminous stars.
Observations of this transition, compared with detailed numerical simulations, are required to confirm these findings and advance our understanding of the winds in CWBs.

The region of parameter space where radiative shocks occur due to IC cooling is also  the region with the lowest $\tau_\mathrm{cool}/\tau_\mathrm{eq}$ ratio, and so care must be taken that the single-fluid assumption is appropriate.
For WR\,140 $\tau_\mathrm{eq}$ is only slightly shorter than the cooling time in the WR wind, and so the electrons may never reach the adiabatic post-shock temperature.
This result is very sensitive to the wind velocity and and to the degree to which the WR wind is decelerated by the O-star radiation field.
Because the O star in WR\,140 has such a weak wind the wind-collision region is very close to the star, deep in the wind acceleration zone, and so the wind velocity is not very large and $\tau_\mathrm{eq}$ is much shorter than the other relevant timescales.

Applying our results to the CWB system WR\,140, we find that the inclusion of IC cooling can potentially resolve difficulties in modelling two observations, namely that the shocks are radiative near periastron and that the hard X-ray emission decreases around periastron (even after considering absorption effects).
2D and 3D simulations are presented that show the transition between radiative and adiabatic shocks occurs very much as predicted from simple theory.
For the specific case of WR\,140, our 3D simulations show that the hard X-ray emission begins to deviate from the adiabatic result around 100 days before periastron (because the shocks enter the wind acceleration region), and decreases dramatically about 10 days before periastron.
This decrease occurs as shocks become radiative through IC cooling, and is consistent with observations of periastron passages of WR\,140 \citep{PolCorSte21}, providing strong observational support for including IC cooling in numerical simulations.
A deeper investigation of this system could strongly confirm (or challenge) our claims.

Moreover, IC cooling in CWBs allows for the novel possibility to break the degeneracy between clumping factor and mass-loss rate of massive stars via independent determination of the wind momentum ratio, $\eta$, in the wind-collision region.
Therefore, detailed observational tests concerning the X-ray temporal evolution in especially CWBs in the Magellanic Clouds \citep{GagFehSav12}, for which the luminosity of the binary stars can be more precisely determined, could provide much needed advancements in the determination of the final properties of massive stars.

IC cooling is not difficult to implement in numerical simulations, and its decisive role in determining the X-ray lightcurve and shock thermodynamics in CWB systems demands that it be included, especially for the close binaries with radiative shocks.
The inclusion of IC cooling and magnetic fields brings a significant leap forward in the predictive power of 3D simulations of CWB systems.
This will be valuable for predicting both thermal and non-thermal radiation from such systems for the next generation of X-ray and $\gamma$-ray observatories.
Careful comparison of models with observation can also yield better constraints on the mass-loss rates of the stars in CWBs, providing important empirical data for stellar evolution calculations.
Mass loss is an important process in determining the final period distribution of the post-supernova binary systems, which in turn is important for interpreting the population of gravitational wave sources from merging compact objects.

\section*{Acknowledgements}
We are grateful to Julian Pittard for the idea to investigate the X-ray emission of WR\,140, for sharing the cooling tables, and for pointing out the 2005 paper where IC cooling was implemented.
TJ is grateful to A.M.T.~Pollock and M.F.~Corcoran for sharing the X-ray data from \citet{PolCorSte21} shown in Fig.~\ref{fig:wr140-3d-mide-xray}.
The authors are grateful to the referees for insightful reviews which has improved the manuscript, and to A.M.T.~Pollock for pointing us towards an error in the equilibration timescale in an earlier draft.
JM acknowledges support from a Royal Society-Science Foundation Ireland University Research Fellowship.
AM acknowledges support from a Royal Society Research Fellows Enhancement Award 2021.
This work was supported by an Irish Research Council (IRC) Starting Laureate Award.
RB acknowledges funding from the Irish Research Council under the Government of Ireland Postdoctoral Fellowship program.
This research made use of VisIt \citep{VisIt}, Astropy \citep{astropy:2018},  Numpy \citep{HarMilVan20}, matplotlib \citep{Hun07} and yt \citep{TurSmiOis11}.

\section*{Data Availability}

The IC cooling and orbital motion modules will be included in the next public release of \textsc{pion}.
An advance copy can be made available following reasonable request.



\bibliographystyle{mnras}
\bibliography{refs} 

\begin{thebibliography}{}
\makeatletter
\relax
\def\mn@urlcharsother{\let\do\@makeother \do\$\do\&\do\#\do\^\do\_\do\%\do\~}
\def\mn@doi{\begingroup\mn@urlcharsother \@ifnextchar [ {\mn@doi@}
  {\mn@doi@[]}}
\def\mn@doi@[#1]#2{\def\@tempa{#1}\ifx\@tempa\@empty \href
  {http://dx.doi.org/#2} {doi:#2}\else \href {http://dx.doi.org/#2} {#1}\fi
  \endgroup}
\def\mn@eprint#1#2{\mn@eprint@#1:#2::\@nil}
\def\mn@eprint@arXiv#1{\href {http://arxiv.org/abs/#1} {{\tt arXiv:#1}}}
\def\mn@eprint@dblp#1{\href {http://dblp.uni-trier.de/rec/bibtex/#1.xml}
  {dblp:#1}}
\def\mn@eprint@#1:#2:#3:#4\@nil{\def\@tempa {#1}\def\@tempb {#2}\def\@tempc
  {#3}\ifx \@tempc \@empty \let \@tempc \@tempb \let \@tempb \@tempa \fi \ifx
  \@tempb \@empty \def\@tempb {arXiv}\fi \@ifundefined
  {mn@eprint@\@tempb}{\@tempb:\@tempc}{\expandafter \expandafter \csname
  mn@eprint@\@tempb\endcsname \expandafter{\@tempc}}}

\bibitem[\protect\citeauthoryear{{Abdo} et~al.,}{{Abdo}
  et~al.}{2010}]{FERMI_2010_EtaCar}
{Abdo} A.~A.,  et~al., 2010, \mn@doi [\apj]
  {10.1088/0004-637X/723/1/64910.48550/arXiv.1008.3235}, \href
  {https://ui.adsabs.harvard.edu/abs/2010ApJ...723..649A} {723, 649}

\bibitem[\protect\citeauthoryear{{Actis}, {Agnetta}, {Aharonian},
  {Akhperjanian}, {Aleksi{\'c}}  \& et al.}{{Actis} et~al.}{2011}]{ActAgnAha11}
{Actis} M.,  {Agnetta} G.,  {Aharonian} F.,  {Akhperjanian} A.,  {Aleksi{\'c}}
  J.,   et al. 2011, \mn@doi [Experimental Astronomy]
  {10.1007/s10686-011-9247-0}, \href
  {http://adsabs.harvard.edu/abs/2011ExA....32..193A} {32, 193}

\bibitem[\protect\citeauthoryear{{Arnaud}}{{Arnaud}}{1996}]{Arn96}
{Arnaud} K.~A.,  1996, in {Jacoby} G.~H.,  {Barnes} J.,  eds,  Astronomical
  Society of the Pacific Conference Series Vol. 101, Astronomical Data Analysis
  Software and Systems V. p.~17

\bibitem[\protect\citeauthoryear{{Astropy Collaboration} et~al.,}{{Astropy
  Collaboration} et~al.}{2018}]{astropy:2018}
{Astropy Collaboration} et~al., 2018, \mn@doi [\aj] {10.3847/1538-3881/aabc4f},
  \href {https://ui.adsabs.harvard.edu/abs/2018AJ....156..123A} {156, 123}

\bibitem[\protect\citeauthoryear{{Bell}, {Schure}  \& {Reville}}{{Bell}
  et~al.}{2011}]{Bel11}
{Bell} A.~R.,  {Schure} K.~M.,   {Reville} B.,  2011, \mn@doi [\mnras]
  {10.1111/j.1365-2966.2011.19571.x}, \href
  {https://ui.adsabs.harvard.edu/abs/2011MNRAS.418.1208B} {418, 1208}

\bibitem[\protect\citeauthoryear{{Birkinshaw}}{{Birkinshaw}}{1999}]{Bir99}
{Birkinshaw} M.,  1999, \mn@doi [\physrep] {10.1016/S0370-1573(98)00080-5},
  \href {https://ui.adsabs.harvard.edu/abs/1999PhR...310...97B} {310, 97}

\bibitem[\protect\citeauthoryear{{Brands} et~al.,}{{Brands}
  et~al.}{2022}]{BraDeKBes22}
{Brands} S.~A.,  et~al., 2022, \mn@doi [\aap] {10.1051/0004-6361/202142742},
  \href {https://ui.adsabs.harvard.edu/abs/2022A&A...663A..36B} {663, A36}

\bibitem[\protect\citeauthoryear{{Cherchneff}}{{Cherchneff}}{2015}]{Che15}
{Cherchneff} I.,  2015, in {Hamann} W.-R.,  {Sander} A.,   {Todt} H.,  eds,
  Wolf-Rayet Stars. pp 269--274

\bibitem[\protect\citeauthoryear{{Cherepashchuk}}{{Cherepashchuk}}{1976}]{Che76}
{Cherepashchuk} A.~M.,  1976, Soviet Astronomy Letters, \href
  {https://ui.adsabs.harvard.edu/abs/1976SvAL....2..138C} {2, 138}

\bibitem[\protect\citeauthoryear{Childs et~al.,}{Childs et~al.}{2012}]{VisIt}
Childs H.,  et~al., 2012, in , {High Performance Visualization--Enabling
  Extreme-Scale Scientific Insight}.
pp 357--372

\bibitem[\protect\citeauthoryear{{Crowther}, {Broos}, {Townsley}, {Pollock},
  {Tehrani}  \& {Gagn{\'e}}}{{Crowther} et~al.}{2022}]{CroBroTow22}
{Crowther} P.~A.,  {Broos} P.~S.,  {Townsley} L.~K.,  {Pollock} A. M.~T.,
  {Tehrani} K.~A.,   {Gagn{\'e}} M.,  2022, \mn@doi [\mnras]
  {10.1093/mnras/stac1952}, \href
  {https://ui.adsabs.harvard.edu/abs/2022MNRAS.515.4130C} {515, 4130}

\bibitem[\protect\citeauthoryear{{Dedner}, {Kemm}, {Kr{\"o}ner}, {Munz},
  {Schnitzer}  \& {Wesenberg}}{{Dedner} et~al.}{2002}]{DedKemKro02}
{Dedner} A.,  {Kemm} F.,  {Kr{\"o}ner} D.,  {Munz} C.-D.,  {Schnitzer} T.,
  {Wesenberg} M.,  2002, \mn@doi [Journal of Computational Physics]
  {10.1006/jcph.2001.6961}, \href
  {http://adsabs.harvard.edu/abs/2002JCoPh.175..645D} {175, 645}

\bibitem[\protect\citeauthoryear{{Derigs}, {Winters}, {Gassner}, {Walch}  \&
  {Bohm}}{{Derigs} et~al.}{2018}]{DerWinGas18}
{Derigs} D.,  {Winters} A.~R.,  {Gassner} G.~J.,  {Walch} S.,   {Bohm} M.,
  2018, \mn@doi [Journal of Computational Physics] {10.1016/j.jcp.2018.03.002},
  \href {https://ui.adsabs.harvard.edu/abs/2018JCoPh.364..420D} {364, 420}

\bibitem[\protect\citeauthoryear{{Dougherty}, {Beasley}, {Claussen}, {Zauderer}
   \& {Bolingbroke}}{{Dougherty} et~al.}{2005}]{DouBeaCla05}
{Dougherty} S.~M.,  {Beasley} A.~J.,  {Claussen} M.~J.,  {Zauderer} B.~A.,
  {Bolingbroke} N.~J.,  2005, \mn@doi [\apj] {10.1086/428494}, \href
  {https://ui.adsabs.harvard.edu/abs/2005ApJ...623..447D} {623, 447}

\bibitem[\protect\citeauthoryear{{Driessen}, {Sundqvist}  \&
  {Dagore}}{{Driessen} et~al.}{2022}]{DriSunDag22}
{Driessen} F.~A.,  {Sundqvist} J.~O.,   {Dagore} A.,  2022, \mn@doi [\aap]
  {10.1051/0004-6361/202142844}, \href
  {https://ui.adsabs.harvard.edu/abs/2022A&A...663A..40D} {663, A40}

\bibitem[\protect\citeauthoryear{{Drury}}{{Drury}}{1983}]{Dru83}
{Drury} L.~O.,  1983, \mn@doi [Reports on Progress in Physics]
  {10.1088/0034-4885/46/8/002}, \href
  {http://adsabs.harvard.edu/abs/1983RPPh...46..973D} {46, 973}

\bibitem[\protect\citeauthoryear{{Eatson}, {Pittard}  \& {Van Loo}}{{Eatson}
  et~al.}{2022a}]{EatPitVan22a}
{Eatson} J.~W.,  {Pittard} J.~M.,   {Van Loo} S.,  2022a, \mn@doi [\mnras]
  {10.1093/mnras/stac2617}, \href
  {https://ui.adsabs.harvard.edu/abs/2022MNRAS.516.6132E} {516, 6132}

\bibitem[\protect\citeauthoryear{{Eatson}, {Pittard}  \& {Van Loo}}{{Eatson}
  et~al.}{2022b}]{EatPitVan22b}
{Eatson} J.~W.,  {Pittard} J.~M.,   {Van Loo} S.,  2022b, \mn@doi [\mnras]
  {10.1093/mnras/stac3000}, \href
  {https://ui.adsabs.harvard.edu/abs/2022MNRAS.517.4705E} {517, 4705}

\bibitem[\protect\citeauthoryear{{Eichler} \& {Usov}}{{Eichler} \&
  {Usov}}{1993}]{EicUso93}
{Eichler} D.,  {Usov} V.,  1993, \mn@doi [\apj] {10.1086/172130}, \href
  {https://ui.adsabs.harvard.edu/abs/1993ApJ...402..271E} {402, 271}

\bibitem[\protect\citeauthoryear{{Eldridge} \& {Stanway}}{{Eldridge} \&
  {Stanway}}{2022}]{EldSta22}
{Eldridge} J.~J.,  {Stanway} E.~R.,  2022, \mn@doi [\araa]
  {10.1146/annurev-astro-052920-100646}, \href
  {https://ui.adsabs.harvard.edu/abs/2022ARA&A..60..455E} {60, 455}

\bibitem[\protect\citeauthoryear{{Faucher-Gigu{\`e}re} \&
  {Quataert}}{{Faucher-Gigu{\`e}re} \& {Quataert}}{2012}]{FauQua12}
{Faucher-Gigu{\`e}re} C.-A.,  {Quataert} E.,  2012, \mn@doi [\mnras]
  {10.1111/j.1365-2966.2012.21512.x}, \href
  {https://ui.adsabs.harvard.edu/abs/2012MNRAS.425..605F} {425, 605}

\bibitem[\protect\citeauthoryear{{Foellmi}, {Moffat}  \& {Guerrero}}{{Foellmi}
  et~al.}{2003a}]{FoeMofGue03a}
{Foellmi} C.,  {Moffat} A.~F.~J.,   {Guerrero} M.~A.,  2003a, \mn@doi [\mnras]
  {10.1046/j.1365-8711.2003.06052.x}, \href
  {https://ui.adsabs.harvard.edu/abs/2003MNRAS.338..360F} {338, 360}

\bibitem[\protect\citeauthoryear{{Foellmi}, {Moffat}  \& {Guerrero}}{{Foellmi}
  et~al.}{2003b}]{FoeMofGue03b}
{Foellmi} C.,  {Moffat} A.~F.~J.,   {Guerrero} M.~A.,  2003b, \mn@doi [\mnras]
  {10.1046/j.1365-8711.2003.06161.x}, \href
  {https://ui.adsabs.harvard.edu/abs/2003MNRAS.338.1025F} {338, 1025}

\bibitem[\protect\citeauthoryear{{Frank}, {Jones}, {Ryu}  \& {Gaalaas}}{{Frank}
  et~al.}{1996}]{FraJonRyu96}
{Frank} A.,  {Jones} T.~W.,  {Ryu} D.,   {Gaalaas} J.~B.,  1996, \mn@doi [\apj]
  {10.1086/177009}, \href
  {https://ui.adsabs.harvard.edu/abs/1996ApJ...460..777F} {460, 777}

\bibitem[\protect\citeauthoryear{{Gagn{\'e}}, {Fehon}, {Savoy}, {Cartagena},
  {Cohen}  \& {Owocki}}{{Gagn{\'e}} et~al.}{2012}]{GagFehSav12}
{Gagn{\'e}} M.,  {Fehon} G.,  {Savoy} M.~R.,  {Cartagena} C.~A.,  {Cohen}
  D.~H.,   {Owocki} S.~P.,  2012, in {Drissen} L.,  {Robert} C.,  {St-Louis}
  N.,   {Moffat} A.~F.~J.,  eds,  Astronomical Society of the Pacific
  Conference Series Vol. 465, Proceedings of a Scientific Meeting in Honor of
  Anthony F. J. Moffat. p.~301 (\mn@eprint {arXiv} {1205.3510}),
  \mn@doi{10.48550/arXiv.1205.3510}

\bibitem[\protect\citeauthoryear{{Grassitelli}, {Chen{\'e}}, {Sanyal},
  {Langer}, {St-Louis}, {Bestenlehner}  \& {Fossati}}{{Grassitelli}
  et~al.}{2016}]{GraCheSan16}
{Grassitelli} L.,  {Chen{\'e}} A.~N.,  {Sanyal} D.,  {Langer} N.,  {St-Louis}
  N.,  {Bestenlehner} J.~M.,   {Fossati} L.,  2016, \mn@doi [\aap]
  {10.1051/0004-6361/201527873}, \href
  {https://ui.adsabs.harvard.edu/abs/2016A&A...590A..12G} {590, A12}

\bibitem[\protect\citeauthoryear{{Grassitelli}, {Langer}, {Grin}, {Mackey},
  {Bestenlehner}  \& {Gr{\"a}fener}}{{Grassitelli} et~al.}{2018}]{GraLanGri18}
{Grassitelli} L.,  {Langer} N.,  {Grin} N.~J.,  {Mackey} J.,  {Bestenlehner}
  J.~M.,   {Gr{\"a}fener} G.,  2018, \mn@doi [\aap]
  {10.1051/0004-6361/201731542}, \href
  {https://ui.adsabs.harvard.edu/abs/2018A&A...614A..86G} {614, A86}

\bibitem[\protect\citeauthoryear{{Green}, {Mackey}, {Haworth}, {Gvaramadze}  \&
  {Duffy}}{{Green} et~al.}{2019}]{GreMacHaw19}
{Green} S.,  {Mackey} J.,  {Haworth} T.~J.,  {Gvaramadze} V.~V.,   {Duffy} P.,
  2019, \mn@doi [\aap] {10.1051/0004-6361/201834832}, \href
  {https://ui.adsabs.harvard.edu/abs/2019A&A...625A...4G} {625, A4}

\bibitem[\protect\citeauthoryear{{Green}, {Mackey}, {Kavanagh}, {Haworth},
  {Moutzouri}  \& {Gvaramadze}}{{Green} et~al.}{2022}]{GreMacKav22}
{Green} S.,  {Mackey} J.,  {Kavanagh} P.,  {Haworth} T.~J.,  {Moutzouri} M.,
  {Gvaramadze} V.~V.,  2022, \mn@doi [\aap] {10.1051/0004-6361/202243531},
  \href {https://ui.adsabs.harvard.edu/abs/2022A&A...665A..35G} {665, A35}

\bibitem[\protect\citeauthoryear{{H.E.S.S. Collaboration} et~al.,}{{H.E.S.S.
  Collaboration} et~al.}{2020}]{HESS20_EtaCar}
{H.E.S.S. Collaboration} et~al., 2020, \mn@doi [\aap]
  {10.1051/0004-6361/201936761}, \href
  {https://ui.adsabs.harvard.edu/abs/2020A&A...635A.167H} {635, A167}

\bibitem[\protect\citeauthoryear{{Hamaguchi} et~al.,}{{Hamaguchi}
  et~al.}{2018}]{2018NatAs...2..731H}
{Hamaguchi} K.,  et~al., 2018, \mn@doi [Nature Astronomy]
  {10.1038/s41550-018-0505-1}, \href
  {https://ui.adsabs.harvard.edu/abs/2018NatAs...2..731H} {2, 731}

\bibitem[\protect\citeauthoryear{{Han} et~al.,}{{Han}
  et~al.}{2020}]{HanTutLau20}
{Han} Y.,  et~al., 2020, \mn@doi [\mnras] {10.1093/mnras/staa2349}, \href
  {https://ui.adsabs.harvard.edu/abs/2020MNRAS.498.5604H} {498, 5604}

\bibitem[\protect\citeauthoryear{{Han}, {Tuthill}, {Lau}  \& {Soulain}}{{Han}
  et~al.}{2022}]{HanTutLau22}
{Han} Y.,  {Tuthill} P.~G.,  {Lau} R.~M.,   {Soulain} A.,  2022, \mn@doi [\nat]
  {10.1038/s41586-022-05155-5}, \href
  {https://ui.adsabs.harvard.edu/abs/2022Natur.610..269H} {610, 269}

\bibitem[\protect\citeauthoryear{Harris et~al.,}{Harris
  et~al.}{2020}]{HarMilVan20}
Harris C.~R.,  et~al., 2020, \mn@doi [Nature] {10.1038/s41586-020-2649-2}, 585,
  357

\bibitem[\protect\citeauthoryear{{Hopkins} et~al.,}{{Hopkins}
  et~al.}{2018}]{HopWetKer18}
{Hopkins} P.~F.,  et~al., 2018, \mn@doi [\mnras] {10.1093/mnras/sty1690}, \href
  {https://ui.adsabs.harvard.edu/abs/2018MNRAS.480..800H} {480, 800}

\bibitem[\protect\citeauthoryear{Hunter}{Hunter}{2007}]{Hun07}
Hunter J.~D.,  2007, \mn@doi [Computing in Science \& Engineering]
  {10.1109/MCSE.2007.55}, 9, 90

\bibitem[\protect\citeauthoryear{{King}}{{King}}{2003}]{Kin03}
{King} A.,  2003, \mn@doi [\apjl] {10.1086/379143}, \href
  {https://ui.adsabs.harvard.edu/abs/2003ApJ...596L..27K} {596, L27}

\bibitem[\protect\citeauthoryear{{Lamberts}, {Dubus}, {Lesur}  \&
  {Fromang}}{{Lamberts} et~al.}{2012}]{LamDubLes12}
{Lamberts} A.,  {Dubus} G.,  {Lesur} G.,   {Fromang} S.,  2012, \mn@doi [\aap]
  {10.1051/0004-6361/201219006}, \href
  {https://ui.adsabs.harvard.edu/abs/2012A&A...546A..60L} {546, A60}

\bibitem[\protect\citeauthoryear{{Lamberts} et~al.,}{{Lamberts}
  et~al.}{2017}]{LamMilLie17}
{Lamberts} A.,  et~al., 2017, \mn@doi [\mnras] {10.1093/mnras/stx588}, \href
  {https://ui.adsabs.harvard.edu/abs/2017MNRAS.468.2655L} {468, 2655}

\bibitem[\protect\citeauthoryear{{Langer}}{{Langer}}{2012}]{Lan12}
{Langer} N.,  2012, \mn@doi [\araa] {10.1146/annurev-astro-081811-125534},
  \href {http://adsabs.harvard.edu/abs/2012ARA%26A..50..107L} {50, 107}

\bibitem[\protect\citeauthoryear{{Langer} et~al.,}{{Langer}
  et~al.}{2020}]{LanSchSto20}
{Langer} N.,  et~al., 2020, \mn@doi [\aap] {10.1051/0004-6361/201937375}, \href
  {https://ui.adsabs.harvard.edu/abs/2020A&A...638A..39L} {638, A39}

\bibitem[\protect\citeauthoryear{{Lau}, {Eldridge}, {Hankins}, {Lamberts},
  {Sakon}  \& {Williams}}{{Lau} et~al.}{2020a}]{LauEldHan20}
{Lau} R.~M.,  {Eldridge} J.~J.,  {Hankins} M.~J.,  {Lamberts} A.,  {Sakon} I.,
   {Williams} P.~M.,  2020a, \mn@doi [\apj] {10.3847/1538-4357/ab9cb5}, \href
  {https://ui.adsabs.harvard.edu/abs/2020ApJ...898...74L} {898, 74}

\bibitem[\protect\citeauthoryear{{Lau} et~al.,}{{Lau}
  et~al.}{2020b}]{LauHanHan20}
{Lau} R.~M.,  et~al., 2020b, \mn@doi [\apj] {10.3847/1538-4357/abaab8}, \href
  {https://ui.adsabs.harvard.edu/abs/2020ApJ...900..190L} {900, 190}

\bibitem[\protect\citeauthoryear{{Lau} et~al.,}{{Lau}
  et~al.}{2022}]{LauHanHan22}
{Lau} R.~M.,  et~al., 2022, \mn@doi [Nature Astronomy]
  {10.1038/s41550-022-01812-x}, \href
  {https://ui.adsabs.harvard.edu/abs/2022NatAs...6.1308L} {6, 1308}

\bibitem[\protect\citeauthoryear{{Luo}, {McCray}  \& {Mac Low}}{{Luo}
  et~al.}{1990}]{LuoMcCMac90}
{Luo} D.,  {McCray} R.,   {Mac Low} M.-M.,  1990, \mn@doi [\apj]
  {10.1086/169263}, \href
  {https://ui.adsabs.harvard.edu/abs/1990ApJ...362..267L} {362, 267}

\bibitem[\protect\citeauthoryear{{Mackey}, {Green}, {Moutzouri}, {Haworth},
  {Kavanagh}, {Zargaryan}  \& {Celeste}}{{Mackey} et~al.}{2021}]{MacGreMou21}
{Mackey} J.,  {Green} S.,  {Moutzouri} M.,  {Haworth} T.~J.,  {Kavanagh} R.~D.,
   {Zargaryan} D.,   {Celeste} M.,  2021, \mn@doi [\mnras]
  {10.1093/mnras/stab781}, \href
  {https://ui.adsabs.harvard.edu/abs/2021MNRAS.504..983M} {504, 983}

\bibitem[\protect\citeauthoryear{{Madura} et~al.,}{{Madura}
  et~al.}{2013}]{MadGulOka13}
{Madura} T.~I.,  et~al., 2013, \mn@doi [\mnras] {10.1093/mnras/stt1871}, \href
  {https://ui.adsabs.harvard.edu/abs/2013MNRAS.436.3820M} {436, 3820}

\bibitem[\protect\citeauthoryear{{Mart{\'\i}-Devesa} \&
  {Reimer}}{{Mart{\'\i}-Devesa} \& {Reimer}}{2021}]{MarRei21}
{Mart{\'\i}-Devesa} G.,  {Reimer} O.,  2021, \mn@doi [\aap]
  {10.1051/0004-6361/202140451}, \href
  {https://ui.adsabs.harvard.edu/abs/2021A&A...654A..44M} {654, A44}

\bibitem[\protect\citeauthoryear{{Mart{\'\i}-Devesa}, {Reimer}, {Li}  \&
  {Torres}}{{Mart{\'\i}-Devesa} et~al.}{2020}]{MarReiLi20}
{Mart{\'\i}-Devesa} G.,  {Reimer} O.,  {Li} J.,   {Torres} D.~F.,  2020,
  \mn@doi [\aap] {10.1051/0004-6361/202037462}, \href
  {https://ui.adsabs.harvard.edu/abs/2020A&A...635A.141M} {635, A141}

\bibitem[\protect\citeauthoryear{{Mignone}, {Zanni}, {Tzeferacos}, {van
  Straalen}, {Colella}  \& {Bodo}}{{Mignone} et~al.}{2012}]{MigZanTze12}
{Mignone} A.,  {Zanni} C.,  {Tzeferacos} P.,  {van Straalen} B.,  {Colella} P.,
    {Bodo} G.,  2012, \mn@doi [\apjs] {10.1088/0067-0049/198/1/7}, \href
  {http://adsabs.harvard.edu/abs/2012ApJS..198....7M} {198, 7}

\bibitem[\protect\citeauthoryear{{Miyamoto}, {Sugawara}, {Maeda}, {Ishida},
  {Hamaguchi}, {Corcoran}, {Russell}  \& {Moffat}}{{Miyamoto}
  et~al.}{2022}]{MiySugMae22}
{Miyamoto} A.,  {Sugawara} Y.,  {Maeda} Y.,  {Ishida} M.,  {Hamaguchi} K.,
  {Corcoran} M.,  {Russell} C. M.~P.,   {Moffat} A. F.~J.,  2022, \mn@doi
  [\mnras] {10.1093/mnras/stac1289}, \href
  {https://ui.adsabs.harvard.edu/abs/2022MNRAS.513.6074M} {513, 6074}

\bibitem[\protect\citeauthoryear{{Moens}, {Poniatowski}, {Hennicker},
  {Sundqvist}, {El Mellah}  \& {Kee}}{{Moens} et~al.}{2022}]{MoePonHen22}
{Moens} N.,  {Poniatowski} L.~G.,  {Hennicker} L.,  {Sundqvist} J.~O.,  {El
  Mellah} I.,   {Kee} N.~D.,  2022, \mn@doi [\aap]
  {10.1051/0004-6361/202243451}, \href
  {https://ui.adsabs.harvard.edu/abs/2022A&A...665A..42M} {665, A42}

\bibitem[\protect\citeauthoryear{{Moffat} \& {Corcoran}}{{Moffat} \&
  {Corcoran}}{2009}]{MofCor09}
{Moffat} A.~F.~J.,  {Corcoran} M.~F.,  2009, \mn@doi [\apj]
  {10.1088/0004-637X/707/1/693}, \href
  {https://ui.adsabs.harvard.edu/abs/2009ApJ...707..693M} {707, 693}

\bibitem[\protect\citeauthoryear{{Myasnikov} \& {Zhekov}}{{Myasnikov} \&
  {Zhekov}}{1993}]{MyaZhe93}
{Myasnikov} A.~V.,  {Zhekov} S.~A.,  1993, \mn@doi [\mnras]
  {10.1093/mnras/260.1.221}, \href
  {https://ui.adsabs.harvard.edu/abs/1993MNRAS.260..221M} {260, 221}

\bibitem[\protect\citeauthoryear{{Pittard}}{{Pittard}}{2009}]{Pit09}
{Pittard} J.~M.,  2009, \mn@doi [\mnras] {10.1111/j.1365-2966.2009.14857.x},
  \href {https://ui.adsabs.harvard.edu/abs/2009MNRAS.396.1743P} {396, 1743}

\bibitem[\protect\citeauthoryear{{Pittard} \& {Dougherty}}{{Pittard} \&
  {Dougherty}}{2006}]{PitDou06}
{Pittard} J.~M.,  {Dougherty} S.~M.,  2006, \mn@doi [\mnras]
  {10.1111/j.1365-2966.2006.10888.x}, \href
  {https://ui.adsabs.harvard.edu/abs/2006MNRAS.372..801P} {372, 801}

\bibitem[\protect\citeauthoryear{{Pittard} \& {Parkin}}{{Pittard} \&
  {Parkin}}{2010}]{PitPar10}
{Pittard} J.~M.,  {Parkin} E.~R.,  2010, \mn@doi [\mnras]
  {10.1111/j.1365-2966.2010.15776.x}, \href
  {https://ui.adsabs.harvard.edu/abs/2010MNRAS.403.1657P} {403, 1657}

\bibitem[\protect\citeauthoryear{{Pollock}}{{Pollock}}{1987}]{Pol87}
{Pollock} A.~M.~T.,  1987, \mn@doi [\apj] {10.1086/165539}, \href
  {https://ui.adsabs.harvard.edu/abs/1987ApJ...320..283P} {320, 283}

\bibitem[\protect\citeauthoryear{{Pollock}, {Corcoran}, {Stevens}  \&
  {Williams}}{{Pollock} et~al.}{2005}]{PolCorSte05}
{Pollock} A.~M.~T.,  {Corcoran} M.~F.,  {Stevens} I.~R.,   {Williams} P.~M.,
  2005, \mn@doi [\apj] {10.1086/431193}, \href
  {https://ui.adsabs.harvard.edu/abs/2005ApJ...629..482P} {629, 482}

\bibitem[\protect\citeauthoryear{{Pollock} et~al.,}{{Pollock}
  et~al.}{2021}]{PolCorSte21}
{Pollock} A.~M.~T.,  et~al., 2021, \mn@doi [\apj] {10.3847/1538-4357/ac2430},
  \href {https://ui.adsabs.harvard.edu/abs/2021ApJ...923..191P} {923, 191}

\bibitem[\protect\citeauthoryear{{Powell}, {Roe}, {Linde}, {Gombosi}  \& {de
  Zeeuw}}{{Powell} et~al.}{1999}]{PowRoeLin99}
{Powell} K.,  {Roe} P.,  {Linde} T.,  {Gombosi} T.,   {de Zeeuw} D.,  1999,
  \mn@doi [Journal of Computational Physics] {10.1006/jcph.1999.6299}, \href
  {http://adsabs.harvard.edu/abs/1999JCoPh.154..284P} {154, 284}

\bibitem[\protect\citeauthoryear{{Pshirkov}}{{Pshirkov}}{2016}]{Psh16}
{Pshirkov} M.~S.,  2016, \mn@doi [\mnras] {10.1093/mnrasl/slv205}, \href
  {https://ui.adsabs.harvard.edu/abs/2016MNRAS.457L..99P} {457, L99}

\bibitem[\protect\citeauthoryear{Rauw}{Rauw}{2022}]{Rau22}
Rauw G.,  2022, X-Ray Emission of Massive Stars and Their Winds.
Springer Nature Singapore, Singapore, pp 1--31 (\mn@eprint {arXiv}
  {2203.16842}), \mn@doi{10.1007/978-981-16-4544-0_79-1}

\bibitem[\protect\citeauthoryear{{Reitberger}, {Kissmann}, {Reimer}  \&
  {Reimer}}{{Reitberger} et~al.}{2017}]{ReiKisRei17}
{Reitberger} K.,  {Kissmann} R.,  {Reimer} A.,   {Reimer} O.,  2017, \mn@doi
  [\apj] {10.3847/1538-4357/aa876d}, \href
  {https://ui.adsabs.harvard.edu/abs/2017ApJ...847...40R} {847, 40}

\bibitem[\protect\citeauthoryear{{Richings} \&
  {Faucher-Gigu{\`e}re}}{{Richings} \& {Faucher-Gigu{\`e}re}}{2018}]{RicFau18}
{Richings} A.~J.,  {Faucher-Gigu{\`e}re} C.-A.,  2018, \mn@doi [\mnras]
  {10.1093/mnras/sty1285}, \href
  {https://ui.adsabs.harvard.edu/abs/2018MNRAS.478.3100R} {478, 3100}

\bibitem[\protect\citeauthoryear{{Russell}, {Corcoran}, {Okazaki}, {Madura}  \&
  {Owocki}}{{Russell} et~al.}{2011}]{RusCorOka11}
{Russell} C. M.~P.,  {Corcoran} M.~F.,  {Okazaki} A.~T.,  {Madura} T.~I.,
  {Owocki} S.~P.,  2011, \mn@doi [Bulletin de la Societe Royale des Sciences de
  Liege] {10.48550/arXiv.1110.1692}, \href
  {https://ui.adsabs.harvard.edu/abs/2011BSRSL..80..719R} {80, 719}

\bibitem[\protect\citeauthoryear{{Rybicki} \& {Lightman}}{{Rybicki} \&
  {Lightman}}{1979}]{RybLig79}
{Rybicki} G.~B.,  {Lightman} A.~P.,  1979, {Radiative processes in
  astrophysics}.
John Wiley \& Sons, Inc., New York

\bibitem[\protect\citeauthoryear{{Sander} \& {Vink}}{{Sander} \&
  {Vink}}{2020}]{SanVin20}
{Sander} A. A.~C.,  {Vink} J.~S.,  2020, \mn@doi [\mnras]
  {10.1093/mnras/staa2712}, \href
  {https://ui.adsabs.harvard.edu/abs/2020MNRAS.499..873S} {499, 873}

\bibitem[\protect\citeauthoryear{{Sander}, {Vink}  \& {Hamann}}{{Sander}
  et~al.}{2020}]{SanVinHam20}
{Sander} A. A.~C.,  {Vink} J.~S.,   {Hamann} W.~R.,  2020, \mn@doi [\mnras]
  {10.1093/mnras/stz3064}, \href
  {https://ui.adsabs.harvard.edu/abs/2020MNRAS.491.4406S} {491, 4406}

\bibitem[\protect\citeauthoryear{{Sekiguchi}, {Tsujimoto}, {Kitamoto},
  {Ishida}, {Hamaguchi}, {Mori}  \& {Tsuboi}}{{Sekiguchi}
  et~al.}{2009}]{SekTsuKit09}
{Sekiguchi} A.,  {Tsujimoto} M.,  {Kitamoto} S.,  {Ishida} M.,  {Hamaguchi} K.,
   {Mori} H.,   {Tsuboi} Y.,  2009, \mn@doi [\pasj]
  {10.1093/pasj/61.4.62910.48550/arXiv.0903.3307}, \href
  {https://ui.adsabs.harvard.edu/abs/2009PASJ...61..629S} {61, 629}

\bibitem[\protect\citeauthoryear{{Shenar} et~al.,}{{Shenar}
  et~al.}{2016}]{SheHaiTod16}
{Shenar} T.,  et~al., 2016, \mn@doi [\aap]
  {10.1051/0004-6361/20152791610.48550/arXiv.1604.01022}, \href
  {https://ui.adsabs.harvard.edu/abs/2016A&A...591A..22S} {591, A22}

\bibitem[\protect\citeauthoryear{{Shenar} et~al.,}{{Shenar}
  et~al.}{2019}]{SheSabHai19}
{Shenar} T.,  et~al., 2019, \mn@doi [\aap] {10.1051/0004-6361/201935684}, \href
  {https://ui.adsabs.harvard.edu/abs/2019A&A...627A.151S} {627, A151}

\bibitem[\protect\citeauthoryear{{Smith}}{{Smith}}{2014}]{Smi14}
{Smith} N.,  2014, \mn@doi [\araa] {10.1146/annurev-astro-081913-040025}, \href
  {http://adsabs.harvard.edu/abs/2014ARA%26A..52..487S} {52, 487}

\bibitem[\protect\citeauthoryear{{Soulain}, {Lamberts}, {Millour}, {Tuthill}
  \& {Lau}}{{Soulain} et~al.}{2023}]{SouLamMil23}
{Soulain} A.,  {Lamberts} A.,  {Millour} F.,  {Tuthill} P.,   {Lau} R.~M.,
  2023, \mn@doi [\mnras] {10.1093/mnras/stac2999}, \href
  {https://ui.adsabs.harvard.edu/abs/2023MNRAS.518.3211S} {518, 3211}

\bibitem[\protect\citeauthoryear{{Spitzer}}{{Spitzer}}{1962}]{Spi62}
{Spitzer} L.,  1962, {Physics of Fully Ionized Gases}

\bibitem[\protect\citeauthoryear{{St-Louis}, {Moffat}, {Marchenko}  \&
  {Pittard}}{{St-Louis} et~al.}{2005}]{StLMofMar05}
{St-Louis} N.,  {Moffat} A. F.~J.,  {Marchenko} S.,   {Pittard} J.~M.,  2005,
  \mn@doi [\apj] {10.1086/430585}, \href
  {https://ui.adsabs.harvard.edu/abs/2005ApJ...628..953S} {628, 953}

\bibitem[\protect\citeauthoryear{{Stevens}, {Blondin}  \& {Pollock}}{{Stevens}
  et~al.}{1992}]{SteBloPol92}
{Stevens} I.~R.,  {Blondin} J.~M.,   {Pollock} A.~M.~T.,  1992, \mn@doi [\apj]
  {10.1086/171013}, \href
  {https://ui.adsabs.harvard.edu/abs/1992ApJ...386..265S} {386, 265}

\bibitem[\protect\citeauthoryear{{Sugawara} et~al.,}{{Sugawara}
  et~al.}{2015}]{SugMaeTsu15}
{Sugawara} Y.,  et~al., 2015, \mn@doi [\pasj] {10.1093/pasj/psv099}, \href
  {https://ui.adsabs.harvard.edu/abs/2015PASJ...67..121S} {67, 121}

\bibitem[\protect\citeauthoryear{{Sunyaev} \& {Zeldovich}}{{Sunyaev} \&
  {Zeldovich}}{1972}]{SunZel72}
{Sunyaev} R.~A.,  {Zeldovich} Y.~B.,  1972, Comments on Astrophysics and Space
  Physics, \href {https://ui.adsabs.harvard.edu/abs/1972CoASP...4..173S} {4,
  173}

\bibitem[\protect\citeauthoryear{{Tavani} et~al.,}{{Tavani}
  et~al.}{2009}]{TavSabPia09}
{Tavani} M.,  et~al., 2009, \mn@doi [\apjl] {10.1088/0004-637X/698/2/L142},
  \href {https://ui.adsabs.harvard.edu/abs/2009ApJ...698L.142T} {698, L142}

\bibitem[\protect\citeauthoryear{{Thomas} et~al.,}{{Thomas}
  et~al.}{2021}]{ThoRicEld21}
{Thomas} J.~D.,  et~al., 2021, \mn@doi [\mnras] {10.1093/mnras/stab1181}, \href
  {https://ui.adsabs.harvard.edu/abs/2021MNRAS.504.5221T} {504, 5221}

\bibitem[\protect\citeauthoryear{{Turk}, {Smith}, {Oishi}, {Skory}, {Skillman},
  {Abel}  \& {Norman}}{{Turk} et~al.}{2011}]{TurSmiOis11}
{Turk} M.~J.,  {Smith} B.~D.,  {Oishi} J.~S.,  {Skory} S.,  {Skillman} S.~W.,
  {Abel} T.,   {Norman} M.~L.,  2011, \mn@doi [The Astrophysical Journal
  Supplement Series] {10.1088/0067-0049/192/1/9}, \href
  {https://ui.adsabs.harvard.edu/abs/2011ApJS..192....9T} {192, 9}

\bibitem[\protect\citeauthoryear{{Tuthill}, {Monnier}  \& {Danchi}}{{Tuthill}
  et~al.}{1999}]{TutMonDan99}
{Tuthill} P.~G.,  {Monnier} J.~D.,   {Danchi} W.~C.,  1999, \mn@doi [\nat]
  {10.1038/19033}, \href
  {https://ui.adsabs.harvard.edu/abs/1999Natur.398..487T} {398, 487}

\bibitem[\protect\citeauthoryear{{Usov}}{{Usov}}{1992}]{Uso92}
{Usov} V.~V.,  1992, \mn@doi [\apj] {10.1086/171236}, \href
  {https://ui.adsabs.harvard.edu/abs/1992ApJ...389..635U} {389, 635}

\bibitem[\protect\citeauthoryear{{Vink}}{{Vink}}{2022}]{Vin22}
{Vink} J.~S.,  2022, \mn@doi [\araa] {10.1146/annurev-astro-052920-094949},
  \href {https://ui.adsabs.harvard.edu/abs/2022ARA&A..60..203V} {60, 203}

\bibitem[\protect\citeauthoryear{{White} \& {Chen}}{{White} \&
  {Chen}}{1995}]{WhiChe95}
{White} R.~L.,  {Chen} W.,  1995, in {van der Hucht} K.~A.,  {Williams} P.~M.,
  eds, ~ Vol. 163, Wolf-Rayet Stars: Binaries; Colliding Winds; Evolution.
  p.~438

\bibitem[\protect\citeauthoryear{{White}, {Breuhaus}, {Konno}, {Ohm}, {Reville}
   \& {Hinton}}{{White} et~al.}{2020}]{WhiBreKon20}
{White} R.,  {Breuhaus} M.,  {Konno} R.,  {Ohm} S.,  {Reville} B.,   {Hinton}
  J.~A.,  2020, \mn@doi [\aap] {10.1051/0004-6361/201937031}, \href
  {https://ui.adsabs.harvard.edu/abs/2020A&A...635A.144W} {635, A144}

\bibitem[\protect\citeauthoryear{{Wiersma}, {Schaye}  \& {Smith}}{{Wiersma}
  et~al.}{2009}]{WieSchSmi09}
{Wiersma} R.~P.~C.,  {Schaye} J.,   {Smith} B.~D.,  2009, \mn@doi [\mnras]
  {10.1111/j.1365-2966.2008.14191.x}, \href
  {http://adsabs.harvard.edu/abs/2009MNRAS.393...99W} {393, 99}

\bibitem[\protect\citeauthoryear{{Williams} et~al.,}{{Williams}
  et~al.}{2009}]{WilMarMar09}
{Williams} P.~M.,  et~al., 2009, \mn@doi [\mnras]
  {10.1111/j.1365-2966.2009.14664.x}, \href
  {https://ui.adsabs.harvard.edu/abs/2009MNRAS.395.1749W} {395, 1749}

\bibitem[\protect\citeauthoryear{{Zhekov}}{{Zhekov}}{2021}]{Zhe21}
{Zhekov} S.~A.,  2021, \mn@doi [\mnras] {10.1093/mnras/staa3591}, \href
  {https://ui.adsabs.harvard.edu/abs/2021MNRAS.500.4837Z} {500, 4837}

\bibitem[\protect\citeauthoryear{{Zhekov} \& {Skinner}}{{Zhekov} \&
  {Skinner}}{2000}]{ZheSki00}
{Zhekov} S.~A.,  {Skinner} S.~L.,  2000, \mn@doi [\apj] {10.1086/309176}, \href
  {https://ui.adsabs.harvard.edu/abs/2000ApJ...538..808Z} {538, 808}

\makeatother
\end{thebibliography}



\appendix

\section{Effects of numerical resolution on results near periastron}
\label{app:resolution}

\begin{figure}
    \centering
    \includegraphics[width=0.9\columnwidth]{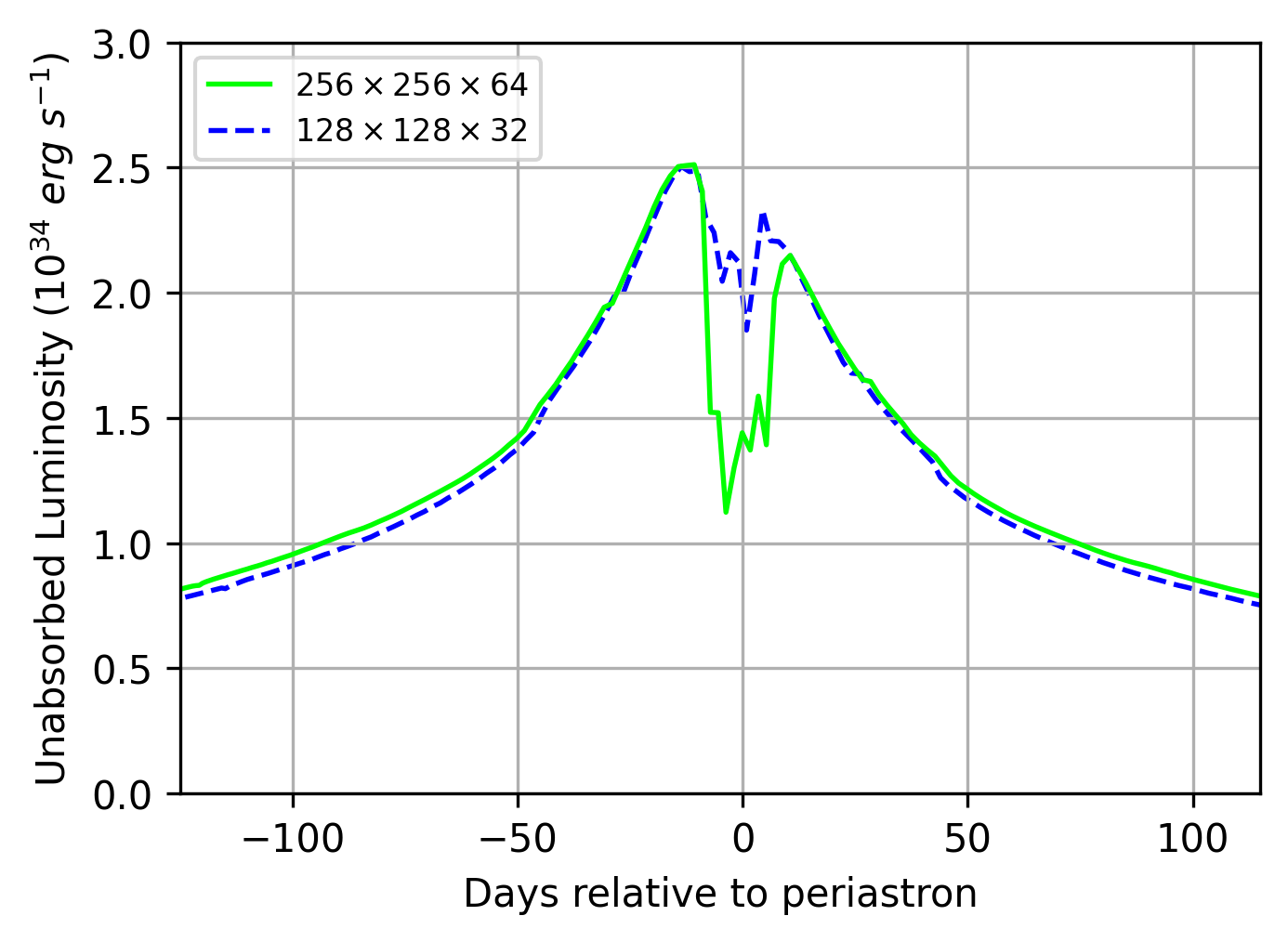}
    \caption[Resolution study for the \texttt{mhd-2} simulation]{Hard X-ray luminosity light-curves for the \texttt{mhd-2} simulation, for resolutions of $128\times128\times32$ and $256\times256\times64$ per grid level.
    The increase in resolution triggers a stronger reduction in X-ray luminosity near periastron when the shocks become radiative.}
    \label{fig:xray-lum-res-study}
\end{figure}

Fig.~\ref{fig:xray-lum-res-study} shows a comparison of the X-ray luminosity curves for the \texttt{mhd-2} simulation for resolutions of $128\times128\times32$ and $256\times256\times64$ cells per level (both simulations have 7 levels of static mesh-refinement).
The simulation with increased resolution shows a greater drop in hard X-ray emission close to periastron.
It is clear that this result has not yet numerically converged, and this can be seen by inspecting the simulation snapshots, which show that the post-shock cooling length has not been spatially resolved.
Further increases in resolution may allow the hard X-ray curve for this \texttt{mhd-2} simulation to drop deeper in luminosity.
We did not have the computational resources to investigate that for this paper but it will be studied in future work.

\bsp	
\label{lastpage}
\end{document}